%% file: pdstack_v6.tex
\documentclass[a4paper,usenatbib]{mnras}

\usepackage{amsmath}	
\usepackage{amssymb}

\usepackage[normalem]{ulem}

\usepackage{graphicx}	

\newcommand{\nest}{\textsc{MultiNest}}

\newcommand\T{\rule{0pt}{2.6ex}}
\newcommand\B{\rule[-1.2ex]{0pt}{0pt}}

\title[Bayesian stacking with confusion noise]{Fully-Bayesian stacking in the presence of confusion}

\author[]{
Song Chen,$^{1}$\thanks{E-mail: phychensong@gmail.com}
Jonathan T.~L.~Zwart$^{1,2}$
and
Mario G.~Santos$^{1,3}$
\\
$^{1}$Department of Physics \& Astronomy, University of the Western
Cape, Bellville, Cape Town 7535, South Africa\\
$^{2}$Astrophysics, Cosmology \& Gravity Centre, Department of Astronomy, University of Cape Town, Rondebosch 7701, South Africa\\
$^{3}$SKA South Africa, 3rd Floor, The Park, Park Road, Pinelands 7405, South Africa\\
}

\date{Accepted XXX. Received YYY; in original form \today}

\pubyear{2017}

\begin{document}
\label{firstpage}
\pagerange{\pageref{firstpage}--\pageref{lastpage}}
\maketitle

\input{abstract}



\input{introduction}

\input{theory_v2}

\input{method}

\input{testing}
\input{simulations}
\input{results}

\input{discussion}
\input{conclusions}

\input{acks}




\bibliographystyle{mnras}
\bibliography{pdstack} 








\bsp	
\label{lastpage}
\end{document}

%% file: abstract.tex
\begin{abstract}
  Multi-wavelength astronomical studies brings a wealth of science
  within reach.  One way to achieve a cross-wavelength analysis is via
  `stacking', i.e.~combining precise positional information from an
  image at one wavelength with data from one at another wavelength in
  order to extract source-flux distributions and other derived
  quantities. For the first time we extend stacking to include the effects of
  confusion.  We develop our algorithm in a fully Bayesian framework and
  apply it to the Square Kilometre Array Design Study (SKADS)
  simulation in order to extract galaxy number counts. Previous
  studies have shown that recovered source counts are highly biased high when
  confusion is non-negligible. With this new method, source counts are
  returned correctly.  We also describe a novel estimator for
  quantifying the impact of confusion on stacking analyses.  This
  method is an essential step in exploiting scientific return for
  upcoming deep radio surveys, e.g.~MIGHTEE on MeerKAT.
\end{abstract}


%% file: introduction.tex
\section{Introduction}
\label{sec:intro}

Measurements of radio-source counts provided some of the earliest
tests of cosmology \citep{Ryle1961,RyleClarke, Longair1966}. Later on,
it turned out that the evolution of radio sources is too strong to
draw robust conclusions about the cosmological model describing the
Universe.  Today, source counts can be used to identify new
extragalactic populations and study galaxy evolution (including star
formation rates and luminosity functions) when combined with redshift
information from panchromatic ancillary surveys.

In 1957, Scheuer \citep{scheuer57} first showed that using a
`probability of deflection', or $P(D)$, analysis allows us to
statistically measure the differential number counts from an image
bearing strong confusion.  \footnote{We will quantify the limitations
  of using a $P(D)$ analysis jointly with stacking analyses in
  Sec.\ref{sec:conf_eff}}.  The idea of this method is to extract
source-count information from the histogram of the pixel-fluxes of the
image.  The performance of this method is highly dependent on the level
of confusion.  The first application \citep{hewish61} was to the 4C
data; the $P(D)$ method has most recently been used to measure number
counts down to 1\,$\mu$Jy at 3\,GHz \citep{condon2012, vernstrom2014}.

Over the past decade many authors have also used `stacking' methods to
extract the average properties of populations of extragalactic sources
selected in a different waveband (e.g.~\citealt{farbeyond}, hereafter
Z15, developed the technique proposed by
\citealt{ketron2014}). \cite{jz-skasci} and Z15 give detailed
overviews of the relative merits of the $P(D)$ and stacking analyses
in the age of the Square Kilometre Array (SKA) precursors.

However, even the most sophisticated stacking analyses have not to
date incorporated the effects of the Point Spread Function (PSF) and
confusion. The effect have not until now been dominant
contribution for current surveys, such as the Very Large Array
(VLA), but they must be inescapably accounted for in order to carry
out near-threshold analyses with forthcoming radio-continuum surveys,
most noticeably MIGHTEE \citep{mightee2017}, MWA's GLEAM
\citep{franzen2016} and LOFAR's MSSS \citep{2015A&A...582A.123H}. 
Indeed, outside the radio-continuum community,
\cite{ed2016} have already sounded the alarm about confusion in
\textsc{hi} stacking experiments.

When applied to simulated data based on SKADS \citep{skads2}, the
fully-Bayesian algorithm in Z15 led to source counts that were biased 
by confusing radio sources for a minimum injected source flux of
$0.01\,\mu$Jy, i.e.~the SKADS-S$^3$ limiting flux; 
such a bias arises because the stacking algorithm in Z15 assumes that 
the contribution to the total flux comes from only one stacked galaxy plus noise,
while in this case the flux contributions from other sources are non-negligible.
The stacking counts became consistent with the true, input model only when the confusing source density in the images was decreased by raising the minimum injected source flux.  A
hard confusion `wall' presents itself in any stacking analysis, its
depth being solely a function of the telescope resolution and the
underlying source counts. The former is usually outside our control,
and the latter is at best a target quantity and at worst a nuisance
term ripe for marginalization.


What is therefore needed is a general framework for stacking analyses
in the presence of confusion. In this work, we still cast the problem
in a fully Bayesian framework rather than in a purely
maximum-likelihood one and the Bayesian evidence is used for selecting
between multi-power-law models.

In Section~\ref{sec:stats}, we give a short prescription for the
derivation of the stacking method in the presence of confusion. In
Section~\ref{sec:conf_eff}, we discuss the detectability of any confusion
contribution.  In Section~\ref{sec:sims} we give some details of the simulations
undertaken as well as our Bayesian framework, including the models,
priors and likelihoods.  We present our results in
Section~\ref{sec:results}  and Section~\ref{sec:simul-fit}. 
Finally, we discuss and conclude in Section~\ref{sec:discussion} and \ref{sec:conclusions}.

%% file: theory_v2.tex
\section{Statistical description}
\label{sec:stats}

In this section, we first review the state-of-the-art stacking method
(see Z15), and the $P(D)$ analysis.  After that
we give a full prescription for the derivation of the probability of
finding a stacking pixel of a given flux, giving full consideration to
the effects of confusion.

\subsection{Stacking without confusion}
\label{sec:stats:noconf}

For completion we now briefly review the earlier Bayesian stacking
method. Z15 assumed that the measured flux $S$ for a given galaxy
could be decomposed as,

\begin{equation}
 S= S_1+\mathcal{N}
\end{equation}

\noindent
where $S_1$ is the
underlying intrinsic flux of the stacking galaxy and $\mathcal{N}$ is
the noise of the image which is assumed to follow a Gaussian
distribution with zero mean and variance $\sigma_n^2$.

The probability of finding one stacking galaxy in a pixel $\Omega_{pix}$ with
measured flux in the interval $\left[S,S+{\rm d} S\right]$ is

\begin{eqnarray}\label{eq:std_stk}
 \mathcal{P}_{stk}(S){\rm d} S=\frac{  \int_{-\infty}^{\infty}{\rm d}S_1 \frac{{\rm d}^2N_s}{{\rm d}S{\rm d}\Omega}(S_1)\frac{\Omega_{pix}}{\sigma_n\sqrt{2\pi}}\exp[-\frac{(S-S_1)^2}{2\sigma_n^2}]}{\bar{n}_{s} \Omega_{pix}} {\rm d} S ,
\end{eqnarray}
where $\bar{n}_{s}$ is the mean number of stacking sources per unit solid angle,
\begin{equation}\label{eq:nbarstk}
\bar{n}_{s}=\int_{-\infty}^{\infty}{\rm d}S_1 \frac{{\rm d}^2N_s}{{\rm d}S{\rm d}\Omega}(S_1)
\end{equation}
\noindent
and the lower indices $s$ stand for ``stacking galaxy'' and $stk$ stand for ''stacking method''.

The probability of finding $k$ stacking pixels in the image $\Omega_\text{image}$ with measured flux in the interval
$\left[S,S+{\rm d} S\right]$ obeys a Poisson distribution
with a mean value $\mathcal{P}_\text{stk}(S) \bar{n}_{s} \Omega_\text{image}{\rm d} S$.

The Z15 stacking method works well when the confusion effect is small
compared to the shot noise of the stacking pixel fluxes histogram (see
section~\ref{sec:conf_eff} for a quantitative discussion). However,
for the next-generation deep radio surveys, this will not always be
the case. 
On one hand, the large sky coverage will allow us to have more stacking sources which shrinks the statistical uncertainty.
On the other hand, the confusion noise will increase substantially as surveys reach ever greater depths

\subsection{$P(D)$ analysis}
\label{sec:stats:pofd}

For completion, we now give a short review of the $P(D)$ method. Our
derivation and notations follow the original $P(D)$ analysis paper\citep{scheuer57}.

Considering an observation at a random position on the sky, a point
source of true flux $S$ at position $(x,y)$ \footnote{One is free to
  choose $(\theta, \phi)$ coordinates, and the result will remain
  the same. It is always reasonable to use the flat-sky
  approximation inside the PSF.} is observed with a smoothed flux $X$,

\begin{equation}
 X= B(x,y)S,
\end{equation}

\noindent
where $B(x,y)$ represents the point-spread function (PSF). We
choose a two-dimensional Gaussian PSF function.

The mean number of total sources within the image $R_t$ observed with a smoothed flux $X$
inside the PSF is given by \citep{condon74}:

\begin{equation}\label{eq:Rx}
 R_t\left(X\right){\rm d} X=\iint_{\Omega_\text{map}} \frac{{\rm d}^2N_t}{{\rm d}S{\rm d}\Omega}\left(\frac{X}{B(x,y)}\right) \frac{{\rm d} X}{B(x,y)} {\rm d}x{\rm d}y,
\end{equation}

\noindent
where the lower indices $t$ stand for "total", and the extra factor $B(x,y)$ in the denominator is due to a change
of variables from $X$ to $S$, and $\frac{{\rm d}^2N_t}{{\rm d}S{\rm d}\Omega}$ is the differential
number count of the total galaxy population within the image field, no matter if it is detected or not.

The mean number of sources inside the PSF integrated over all possible
fluxes is:

\begin{equation}
 \bar{n}_t=\int_0^\infty R_t(X) {\rm d}X \;. 
\end{equation}

\noindent
The probability $G(n)$ of having $n$ sources inside the PSF is
a Poisson distribution of mean $\bar{n}_t$,

\begin{eqnarray}
 G(n)&=&\frac{(\bar{n}_t)^n}{n!}\mathrm{e}^{-\bar{n}},
\end{eqnarray}

\noindent
Since the observed flux of each image pixel is a convolution of the
PSF with the flux of galaxies inside inside the PSF plus instrumental noise, 
the probability of an image pixel
having a flux $S$ to $S+{\rm d} S$ (i.e.~ the so-called $P(D)$ where `D' is for
deflection) can be split according to the number of sources inside the
PSF.

\begin{equation}\label{eq:sum}
 \mathcal{P}_D(S){\rm d} S=\sum_{n=0}^\infty G(n) P_n (S){\rm d} S.
\end{equation}

\noindent
where $P_n(S)$ is the probability of having a pixel with flux $S$
knowing there are $n$ galaxies inside the PSF centred at the pixel
position. 

Now, considering the first term in the summation, $n=0$, i.e.~no
galaxy inside the PSF,

\begin{eqnarray}
P_0 (S) =  \mathcal{N} (S),
\end{eqnarray}

\noindent
where $\mathcal{N}(S)$ is the normalized noise probability density
function (PDF), here a Gaussian PDF.
Compared to first term, the second term $n=1$ adds one confusing
galaxy into the PSF:

\begin{eqnarray}
 P_1(S) =\frac{1}{\bar{n}} \int_{-\infty}^{+\infty} {\rm d}X_{1} R_t(X_{1})  \mathcal{N} (S-X_{1}) \;,
\end{eqnarray}
where we extend the integration range to infinity, and set the
differential number count function
$\frac{{\rm d}^2N_t}{{\rm d}S{\rm d}\Omega}$ to be zero for the flux
$S>S_{max}$ or $S<S_{min}$. Similarly, the third term $n=2$ is,

\begin{eqnarray}
P_2(S)&=&\frac{1}{\bar{n}^2 } \int_{-\infty}^{+\infty} {\rm d}X_{1} R_t(X_{1})\\
 &\times&\int_{-\infty}^{+\infty} {\rm d}X_{2} R_t(X_{2})  \mathcal{N} (S-\sum_{j=1}^2 X_{j})\;, \nonumber 
\end{eqnarray}

\noindent
For $n$ galaxies we have

\begin{eqnarray}
 &&P_n(S){\rm d}S=\frac{1}{\bar{n}^n } \int_{-\infty}^{+\infty} {\rm d}X_{1} R_t(X_{1})\\
 &\times&\int_{-\infty}^{+\infty} {\rm d}X_{2} R_t(X_{2})...\int_{-\infty}^{+\infty} {\rm d}X_{n} R_t(X_{n})  \mathcal{N} (S-\sum_{j=1}^n X_{j})\;, \nonumber 
\end{eqnarray}

\noindent
Using the convolution theorem, we see that the Fourier transform of
this term can be simplified as

\begin{eqnarray}
\mathcal{F}\{P_n(S)\}&\equiv&\int_{-\infty}^{\infty} e^{-2\pi i w S}P_n(S) {\rm d}S\\
&=& \exp(-2\pi \sigma^2 w^2)[\frac{r_t(w)}{\bar{n}}]^{n}, \nonumber\\
\end{eqnarray}
where  $r_t(w)$ is the Fourier
transform of $R_t(X)$, i.e.~$r_t(w)\equiv\mathcal{F}\{R_t(X)\}$, and $\exp(-2\pi \sigma^2 w^2)$ is the Fourier transform of the Gaussian noise with standard deviation $\sigma$.



Therefore, the probability of obtaining a total flux $S$ in a PSF is

\begin{eqnarray}\label{eq:pofd}
\mathcal{P}_D(S){\rm d} S&=&\mathcal{F}^{-1}\{\sum G(n) \mathcal{F}\{ P_n(S)\}\}  {\rm d} S  \\
&=&\mathcal{F}^{-1}\{ \exp(-2\pi \sigma^2 w^2)\exp(r_t(w)-\bar{n}) \}  {\rm d} S \nonumber\;. 
 \end{eqnarray}
where we used the Taylor expansion of exponential function. This is the PDF of $P(D)$ analysis used in the previous papers such as \citep{condon2012, vernstrom2014}.

\subsection{Stacking with confusion - low density case}
\label{sec:stats:withconf}

The stacking method outlined in section~\ref{sec:stats:noconf} is based
on the assumption that there is no contribution from other sources
inside the point spread function (PSF). However, this is not always
true: The confusion effect must in general be taken into account.

In this subsection, we consider the confusion effect, where the total
noise is not Gaussian, but rather follows the confusion amplitude
distribution $P(D)$. To distinguish a stacking galaxy from a non-stacking
galaxy, we introduce lower indices to the differential number count
${\rm d}^2N_{s\;\text{OR}\; o}/{\rm d}S{\rm d}\Omega$, with $s$ for stacking
galaxies and $o$ for non-stacking galaxies.

Considering the fact that the pixels we use for the stacking analysis
are always centered at the positions of stacking galaxies, the stacking
galaxy flux is not modified by the PSF (in the limit of small
positional uncertainties). The mean number $R_s$ of stacking galaxies
observed with a total flux $S$ inside the unitary solid angle is

\begin{equation}\label{eq:Rx_stk}
 R_s\left(S\right){\rm d} S\equiv \frac{{\rm d}^2N_s}{{\rm d}S{\rm d}\Omega} \left(S\right) {\rm d}S.
\end{equation}

\noindent
The mean number of stacking galaxies inside the unitary solid angle $\bar{n}_s$ is the same as we defined in Eq.~\ref{eq:nbarstk}


\noindent
The mean number $R_{o}$ of non-stacking galaxies observed with a flux $X$
inside a PSF is, similarly,

\begin{equation}\label{eq:Rx}
 R_{o}\left(X\right){\rm d} X=\iint \frac{{\rm d}^2N_{o}}{{\rm d}S{\rm d}\Omega}\left(\frac{X}{B(x,y)}\right) \frac{{\rm d} X}{B(x,y)} {\rm d}x{\rm d}y,
\end{equation}

\noindent
The mean number of non-stacking galaxies inside a PSF is then an
integral over all fluxes,

\begin{equation}
 \bar{n}_{o}=\int_0^\infty R_{o}(X) {\rm d}X.
\end{equation}

\noindent
we consider the special case that the total
differential number counts is dominated by the non-stacking galaxies,
\textit{not} the stacking galaxies(e.g., the stacked population is sparse).

The conditional probability $\mathcal{P}_\text{stk}(S){\rm d}S$ of a stacking pixel
having a flux $S$ to $S+{\rm d} S$ can be split according to the number of
non-stacking galaxies inside the PSF, i.e.~

\begin{equation}\label{eq:sum_ind}
 \mathcal{P}_\text{stk}(S){\rm d}S=\sum_{n=0}^\infty G^{o}(n) P_n^s (S) {\rm d}S,
\end{equation}
where $G^{o}(n)$ is the probability of having $n$ sources inside the PSF following
a Poisson distribution of mean $\bar{n}_{o}$.

\noindent
Now, considering the $n=0$ case in Eq.~\ref{eq:sum_ind}, there
is only one stacking galaxy with noise,

\begin{eqnarray}
&& P_0^s (S)          \nonumber \\
&=& \int_{-\infty}^{+\infty} \frac{R_s(X_1)}{\bar{n}_s} \mathcal{N} (S-X_1){\rm d}X_1,
\end{eqnarray}

\noindent
where $\frac{R_s(X_1)}{\bar{n}_s}$ gives the probability of a stacking
galaxy have flux $X_1$. Comparing with the previous case, the $n=1$ case adds one non-stacking galaxy:

\begin{eqnarray}
 &&P_1^s(S) =\frac{1}{\bar{n}_s\bar{n}_{o}} \int_{-\infty}^{+\infty} {\rm d}X_{1} R_s(X_{1}) \nonumber\\
 &\times&\int_{-\infty}^{+\infty} {\rm d}X_{1} \mathcal{N} (S-X_{1}-X_{2})R_{o}(X_{2}).
\end{eqnarray}

\noindent
Thus,

\begin{eqnarray}
 &&P_n^s(S)=\frac{1}{\bar{n}_s \bar{n}_{o}^n } \int_{-\infty}^{+\infty} {\rm d}X_{1} R_s(X_{1})\\
 &&\int_{-\infty}^{+\infty} {\rm d}X_{2} R_{o}(X_{2})...\int_{-\infty}^{+\infty} {\rm d}X_{n} R_{o}(X_{n})  \mathcal{N} (S-\sum_{j=1}^{n+1} X_{j})\;, \nonumber 
\end{eqnarray}

\noindent
Similarly to the $P(D)$ derivation, the convolution theorem can
simplify the calculation. Following this route, we find that the Fourier
transform of $G^{o}(n)P_n^s(S)$ is
\begin{eqnarray}
&&\mathcal{F}\{G^{o}(n)P_n^s(S)\}\\
&=&G^{o}(n) p^s_n(w) \nonumber\\
&=&  \frac{(\bar{n}_{o})^n}{n!}\exp(-\bar{n}_{o}) \frac{r_s(w)}{\bar{n}_s} \exp(-2\pi \sigma^2 w^2)[\frac{r_{o}(w)}{\bar{n}_{o}}]^{n}, \nonumber
\end{eqnarray}

\noindent
where $p^s_n(w)\equiv\mathcal{F}\{P^s_n(X)\}$, and $r_{s\;\text{or}\;o}(w)$ is the
Fourier transform of $R_{s\;\text{or}\;o}(X)$,
i.e.~$r_{s\;\text{or}\;o}(w)\equiv\mathcal{F}\{R_{s\;\text{or}\;o}(X)\}$. 
Therefore, the probability of observing a stacking pixel with total flux $S$ is

\begin{eqnarray}\label{eq:stk_confusion_ind}
 \mathcal{P}_\text{stk}(S)&=& \mathcal{F}^{-1}\{ \sum_{n=1}^\infty G^{o}(n) p^s_n(w)  \}   \\
&=& \mathcal{F}^{-1}\{ \frac{r_s(w)}{\bar{n}_s}\exp(r_{o}(w)-\bar{n}_{o}-2\pi \sigma^2 w^2) \}.\nonumber
\end{eqnarray}
This is also the convolution of the stacking probability with the $P(D)$, e.g. replacing the Gaussian noise in the stacking equation by the $P(D)$ equation.


\subsection{Stacking with confusion - high density case}
\label{sec:stats:new_method}

As the density of stacking sources increases, it is getting difficult to measure the non-stacking source differential number count.
In this section, we derive a PDF of having a stacking pixel with a certain flux, using
the total galaxies differential number counts(including the stacking population).

The conditional probability $\mathcal{P}_\text{stk}(S)$ for a stacking
pixel to have a flux $S$ to $S+{\rm d} S$ can be consider as a sum of joint probabilities of having a stacking pixel of flux $S$ contributed from $n_t\in{\Bbb N}$ galaxies inside the PSF.
\begin{equation}\label{eq:sum}
 \mathcal{P}_\text{stk}(S){\rm d}S=\sum_{n_t=1}^{\infty} G'(n) P_{n}' (S){\rm d}S,
\end{equation}
where $G'(n)$ is the conditional
probability of having $n$ total number of galaxies inside a PSF, given the fact that the stacking galaxy is sitting at the center of PSF.
And $P_{n}'(S)$ is the probability of having a stacking pixel of flux $S$, 
knowning that the total number of galaxies inside the PSF is $n$. 
It is reasonable to assume that having $n$ galaxies inside the PSF and their total flux being $S$ are statistical independent.
Thus, the joint probability is simply the product $G'(n) P_{n}'(S)$.

Since we do not consider the clustering effect in this paper, knowning the stacking galaxy is sitting at the center of PSF
do not increase the probability of finding other galaxies inside the same PSF.
Therefore, the probability $G'(n)$ is described by the renormalized Poisson distribution function,
\begin{equation}
   G'(n)=\frac{\bar{n}_t^n}{n!} \frac{ {e}^{-\bar{n}_t}}{1-{e}^{-\bar{n}_t}}\;,
\end{equation}
\noindent
where in the renormalization we exclude the $n=0$ case, and
$\bar{n}_t$ is the mean number of galaxies(including stacking and non-stacking galaxies) inside the PSF,
\begin{equation}
 \bar{n}_t=\int_0^\infty R_t(X) {\rm d}X\;,
\end{equation}
where
\begin{equation}\label{eq:Rt}
 R_t\left(X\right){\rm d} X=\iint \frac{{\rm d}^2N_t}{{\rm d}S{\rm d}\Omega}\left(\frac{X}{B(x,y)}\right) \frac{{\rm d} X}{B(x,y)} {\rm d}x{\rm d}y.
\end{equation}
It is important to point out that $R_t\left(X\right){\rm d} X$ is the average number of galaxies inside the whole PSF, and it has already counted the galaxies sitting at the center of the PSF. 

Next, we consider $P_{n}'(S)$ where $n=1$ case:
\begin{eqnarray}
 P_1' (S)=\int_{-\infty}^{+\infty} \frac{R_s(X_1)}{\bar{n}_s} \mathcal{N} (S-X_1){\rm d}X_1.
\end{eqnarray}

\noindent
For the second case $n=2$,

\begin{eqnarray}
 &&P_2'(S) =\frac{1}{\bar{n}_s\bar{n}_t} \int_{-\infty}^{+\infty} {\rm d}X_{1} R_s(X_{1}) \nonumber\\
 &\times&\int_{-\infty}^{+\infty} {\rm d}X_{2} \mathcal{N} (S-X_{1}-X_{2})R_t(X_{2}),
\end{eqnarray}
\noindent
where the ratio $ R_t\left(X\right){\rm d} X / \bar{n}_t$ gives the probability of having PSF averaged flux(or pixel flux) $X$.

Similarly, for the $n_t=3$ case,

\begin{eqnarray}
 P_3'(S)=\frac{1}{\bar{n}_s\bar{n}_t^2 } \int_{-\infty}^{+\infty} {\rm d}X_{1} R_s(X_{1})&&\nonumber\\
 \int_{-\infty}^{+\infty} {\rm d}X_{2} R_t(X_{2})\int_{-\infty}^{+\infty} {\rm d}X_{3} R_t(X_{3})  \mathcal{N} (S-\sum_j X_{j})\;,&& \nonumber.
\end{eqnarray}

\noindent
Hence for the $n$-th case we have that
\begin{eqnarray}
 &&P_n'(S)=\frac{1}{\bar{n}_s \bar{n}_t^n } \int_{-\infty}^{+\infty} {\rm d}X_{1} R_s(X_{1})\\
 &&\int_{-\infty}^{+\infty} {\rm d}X_{2} R_t(X_{2})...\int_{-\infty}^{+\infty} {\rm d}X_{n} R_t(X_{n})  \mathcal{N} (S-\sum_{j=1}^n X_{j}), \nonumber 
\end{eqnarray}

\noindent
Using the convolution theorem, we see that the Fourier transform $p_{n}'(w)\equiv\mathcal{F}\{P_{n}'(X)\}$ is
\begin{eqnarray}
&& p_n'(w) \\
&=&  \frac{r_s(w)}{\bar{n}_s} \exp(-2\pi \sigma^2 w^2)[\frac{r_t(w)}{\bar{n}_t}]^{n-1}, \nonumber
\end{eqnarray}
\noindent
where, $r_{t}(w)$ is the Fourier transform of $R_{t}(X)$,
i.e.~$r_{t}(w)\equiv\mathcal{F}\{R_{t}(X)\}$. Therefore, the
probability of observing a stacking pixel of total flux $S$ is

\begin{eqnarray}\label{eq:stk_confusion_high}
 \mathcal{P}_\text{stk}(S)&=& \mathcal{F}^{-1}\{ \sum_{n_t=1}^\infty G'(n) p_{n}'(w)  \}   \\
&=& \mathcal{F}^{-1}\{ \frac{\bar{n}_t}{r_t(w)}\frac{r_s(w)}{\bar{n}_s}\frac{\exp(-2\pi \sigma^2 w^2)}{\exp(\bar{n}_t)-1}[\exp(r_t(w))-1] \}.\nonumber
\end{eqnarray}

In section \ref{sec:conf_eff}, we use this PDF to shed light on the bias
in the number counts found by the simulations of Z15, where the stacking sources in the image constitute the
dominant population. 
While, we also use it to estimate the stacking number counts in section \ref{sec:results_high}, where $40\%$ of the total injected galaxies are stacking galaxies. 

%% file: method.tex
\section{Methodology}
\label{sec:recipe}

We would like to constrain the differential number counts of the stacking galaxies close to the confusion limit.  
When the number of stacking galaxies is large, it becomes difficult to measure the non-stacking galaxies differential number counts isolated because masking the stacking galaxies in the map(which usually requires a patch of around $3-\sigma$ of the PSF) will remove most pixels in the image.

When the number density of the stacking galaxies is small compared to the non-stacking galaxies,
the measured total galaxies differential number counts via the $P(D)$
analysis from the full image can be considered as a good approximation of the non-stacking galaxies differential number counts ${\rm d}^2N_{o}/{\rm d}S{\rm d}\Omega \approx {\rm d}^2N_{t}/{\rm d}S{\rm d}\Omega$. 
Alternatively, we can also use the PDF derived in section \ref{sec:stats:new_method}.

For both cases, we first measure the non-stacking(or total) galaxies differential number counts from the full image pixels via the $P(D)$ analysis.  
Then we use the best-fit galaxy differential number counts as an input, and calculate the stacking likelihood via Eq.\ref{eq:stk_confusion_ind} OR Eq.\ref{eq:stk_confusion_high} for different stacking differential number counts models.

Our measurements are two histograms of pixel fluxes . One is for the full image pixels, the other is for the stacking pixels only. 
The statistical uncertainty of each histogram bin is the square root of the total number of pixels belonging to the bin.
Since, the total number of pixels in the image field is larger than the number of stacking pixels in the same image,
the statistical uncertainty of the full image pixel fluxes histogram is smaller than stacking pixel fluxes histogram.
The inferred number counts from the full image pixel fluxes histogram should be tighter constrained as well.

We assume the non-stacking (or total) galaxies differential number counts can be tightly constrained by the full image, 
and the statistical uncertainty of the stacking pixel fluxes histogram is the dominat error.
Under this approximation, we simplify our evaluation by neglecting the uncertainty on the non-stacking (or total) galaxy differential number counts.
In section \ref{sec:simul-fit}, we will fit the total galaxies and stacking galaxies differential number counts simultaneously.

Following \cite{ketron2014} and Z15, we adopt a Poisson likelihood for
the stacking galaxies and $P(D)$ analysis and neglect the correlation between the image pixels\footnote{To consider correlations, it may be easier to use a Gaussian likelihood, and plug in the covariant matrix from simulation as in \cite{vernstrom2014}},

\begin{equation}\label{eq:likelihoodP}
\mathcal{L}=\prod_i \frac{I_i^{k_i}}{k_i !}\exp(-I_i) \;.
\end{equation}
where $k_i$ is observed number of pixels in the stacking pixel
$i^{th}$ histogram bin, and
\begin{equation}
 I_{i}=\bar{N}_\text{stk} \int_{S_i}^{S_i+\delta S} {\rm d}S' \mathcal{P}_\text{stk}(S') \;,
\end{equation}

\noindent
where the stacking PDF $P_\text{stk}(S)$ is shown in Eq.\ref{eq:stk_confusion_ind}, 
and $\bar{N}_\text{stk}$ is the total number of the expected stacking galaxies inside the image field $\Omega_\text{img}$
\begin{equation}
 \bar{N}_\text{stk}=\Omega_\text{img} \int_\text{Smin}^\text{Smax} {\rm d}S \frac{{\rm d}^2N_s}{{\rm d}S{\rm d}\Omega} \left(S\right)
\end{equation}

For the $P(D)$ analysis, we also adopt the Poisson likelihood Eq.\ref{eq:likelihoodP}, except we replace $I_i\to J_i$
\begin{equation}
 J_{i}=N_\text{pix} \int_{S_i}^{S_i+\delta S} {\rm d}S' \mathcal{P}_D(S') \;,
\end{equation}
\noindent
where $P_D(S')$ is the $P(D)$ analysis PDF, i.e. Eq.\ref{eq:pofd}.

\subsection{Algorithm}
\label{sec:algorithm}
The recipe for application of our methods
\footnote{Our stacking likelihood calculation program can be download from \url{https://github.com/phychensong/ConfusIuS}.} 
is summarized as follows:

I. Low density method:
\begin{enumerate}

\item From a radio image, extract the full pixel fluxes histogram.
\item Fit the differential number counts model to
  these noisy data using the $P(D)$ likelihood given in this section.
  Establish the best-fit differential source counts parameters.
\item Associate the best-fit differential number counts model with the \textbf{non-stacking galaxies inside the image}.  
\item Generate a histogram of fluxes extracted only from the positions
  of the source population to be stacked.
\item Fit the stacking galaxies differential number counts model for the
  stacked population using Eq.\ref{eq:stk_confusion_ind}.

\end{enumerate}

II. High density method:
\begin{enumerate}

\item Same as Low density method
\item Same as Low density method
\item Associate the best-fit differential number counts model with the \textbf{entire galaxies inside the image}.  
\item Generate a histogram of fluxes extracted only from the positions
  of the source population to be stacked.
\item Fit the stacking galaxies differential number counts model for the
  stacked population using Eq.\ref{eq:stk_confusion_high}.

\end{enumerate}

\subsection{Sampling}
 
Sampling parameter spaces is often slow, especially when evidence
integrations, which is required for model selection, are carried
out. Nested sampling \citep{skilling04} was introduced specifically
for the purpose of cutting the computational cost of this. However, it
is an inescapable fact that the evidence integrations are exponential
in the number of model parameters, in practice limiting that number to
$\lesssim 100$. The many advantages of nested sampling --- compared to
MCMC methods --- are discussed elsewhere (see e.g.~Z15).

The \textit{de facto} implementation of nested sampling is \nest\
\citep{feroz08,feroz09}, which has a \textsc{python} wrapper
\citep{pymultinest}. We deploy \nest\ on, typically, 48--96
processors, for as many as $10^5$ likelihood calculations in
total.

%% file: testing.tex
\section{Quantifying confusion-induced bias in stacked counts}
\label{sec:conf_eff}

Using the equations derived in Section \ref{sec:stats:new_method}, we
can further study the upward bias in the differential number counts of
a stacking population found by Z15. The convolution theorem allows us
to isolate an effective noise function from the stacking PDF:

\begin{eqnarray}
  \mathcal{P}_\text{stk}(S)&=& \int_{-\infty}^{+\infty} \frac{R_s(S_s)}{\bar{n}_s} \mathcal{N}_{eff} (S-S_s){\rm d}S_s.\nonumber
\end{eqnarray}

\noindent
The density of stacking galaxies used by Z15 was high\footnote{In Z15,
  \textit{all} the galaxies in the image were selected as stacking
  galaxies.}, so we employ Eq.~\ref{eq:stk_confusion_high}.  
The effective-noise function is
\begin{eqnarray}
 & \mathcal{N}_{eff}(S) &\\
 &=&  \mathcal{F}^{-1}\{ \frac{\bar{n}_t}{r_t(w)}\frac{\exp(-2\pi \sigma^2 w^2)}{\exp(\bar{n}_t)-1}[\exp(r_t(w))-1] \}.\nonumber
\end{eqnarray}

\noindent
This effective noise function is essentially the convolution of
Gaussian image noise with the pure confusion effect. In order to see
the difference between the ordinary stacking method and the stacking
method including confusion, we define an excess term $E_{eff}$ that
traces the extra contribution from confusion,

\begin{equation}\label{eq:excess}
 E_{eff}(S){\rm d}S\equiv (\mathcal{N}_{eff}-\mathcal{N})\bar{N}_\text{stk}{\rm d}S,
\end{equation}

\noindent
where $\bar{N}_\text{stk}$ is the total number of stacking pixels
inside the image.  We then compare this excess to the uncertainty of
the measurement, i.e.~the Poisson noise of the bins of the stacking
pixel-flux histogram. This is just the square root of the number of
stacking pixels falling into a certain flux range, and is proportional
to the stacking PDF (i.e.~Eq.\ref{eq:stk_confusion_high}).

\begin{equation}\label{eq:short_noise}
 \mathcal{N}_P(S){\rm d}S=\sqrt{\mathcal{P}_\text{stk}(S) \bar{N}_\text{stk}{\rm d}S}
\end{equation}

\noindent
If the excess is larger than the uncertainty, $\mathcal{N}_p$, of the
measurement, then we can detect this small difference between the
Gaussian image noise and the convolution of the Gaussian image noise
with the confusion effect from a given measurement.  As a result of
this `detection', the fitted stacking-population number counts using
ordinary stacking will be shifted upwards from the true number counts
so as to compensate for this extra confusion contribution.

The same idea can be adopted to describe a $P(D)$ analysis as
well. The excess $E_{eff}^p$ for the $P(D)$ analysis is

\begin{equation}
 E_{eff}'(S){\rm d}S\equiv N_{pix}(\mathcal{P}_D(S)-\mathcal{N}){\rm d} S\;,
\end{equation}

\noindent
where $N_{pix}$ is the total number of pixels in the image. 
The Poisson noise of the bins of the $P(D)$ pixel fluxes histogram is

\begin{equation}\label{eq:short_noise}
 \mathcal{N}_{P}'(S)=\sqrt{\mathcal{P}_D(S) N_{pix} {\rm d}S}
\end{equation}

\noindent
If the the excess $E_{eff}'(S)$ is larger than $\mathcal{N}_p'(S)$,
then we can detect the confusion contribution to the histogram.  In
other words, a $P(D)$ analysis implemented on this data set has
sufficient discriminatory power to extract number counts from the
confusing population.

In order to compare with the previous results from Z15, our
calculation proceeds on the following basis: We set the PSF resolution
to be 6\arcsec\ and the survey area to be $1 \mathrm{deg}^2$. We use
the SKADS-$S^3$ simulation
$\frac{{\rm d}^2N}{{\rm d}S{\rm d}\Omega} \left(S\right)$, and bound
the source flux between $S_{min}=0.1\text{ (and 1.0) }\mu \mathrm{Jy}$
and $S_{max}=85 \mu\mathrm{Jy}$. The stacking sources include all
the simulation sources above the limiting flux $S_{min}$.

\begin{figure}
\includegraphics[width=\columnwidth]{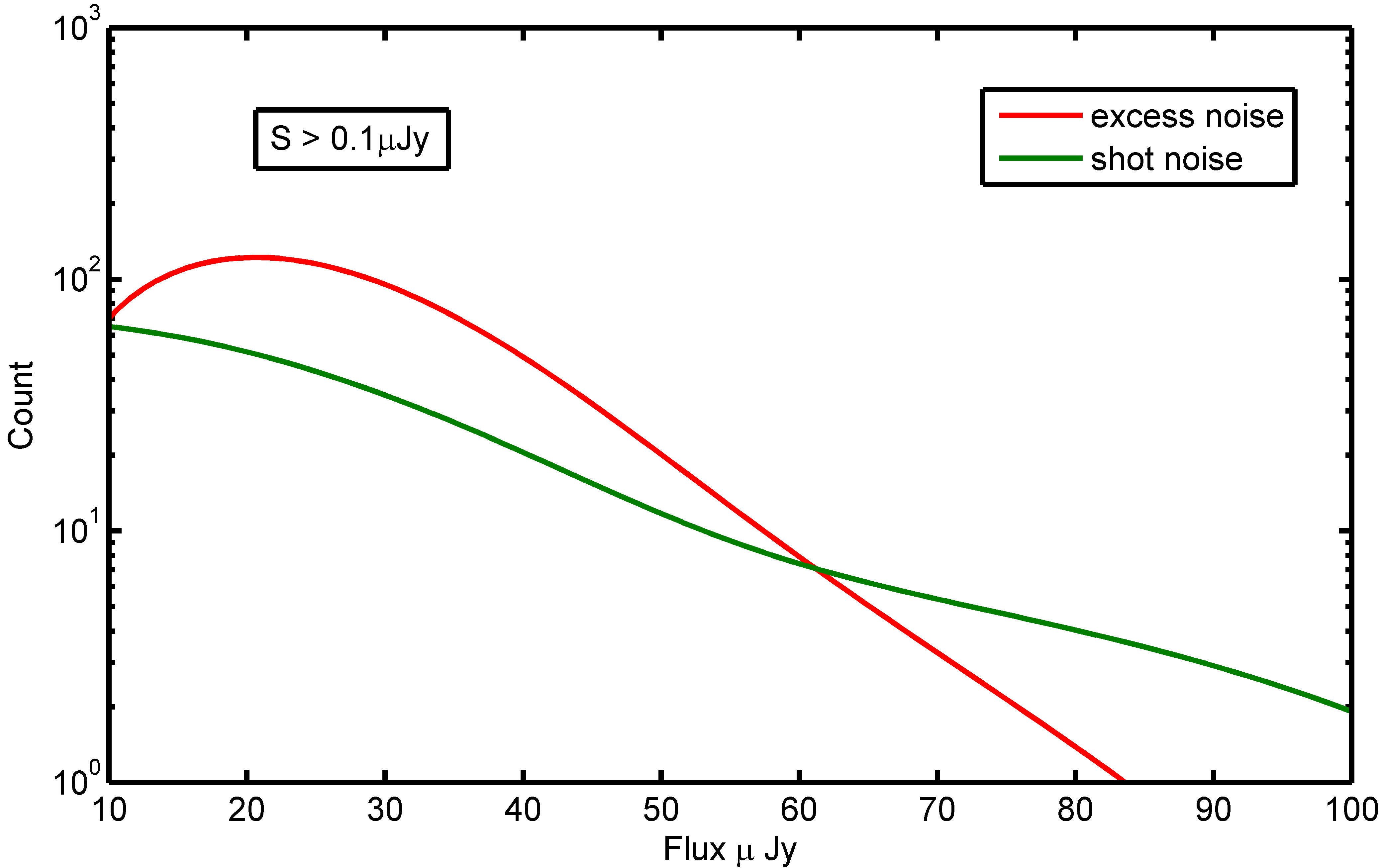}      
\includegraphics[width=\columnwidth]{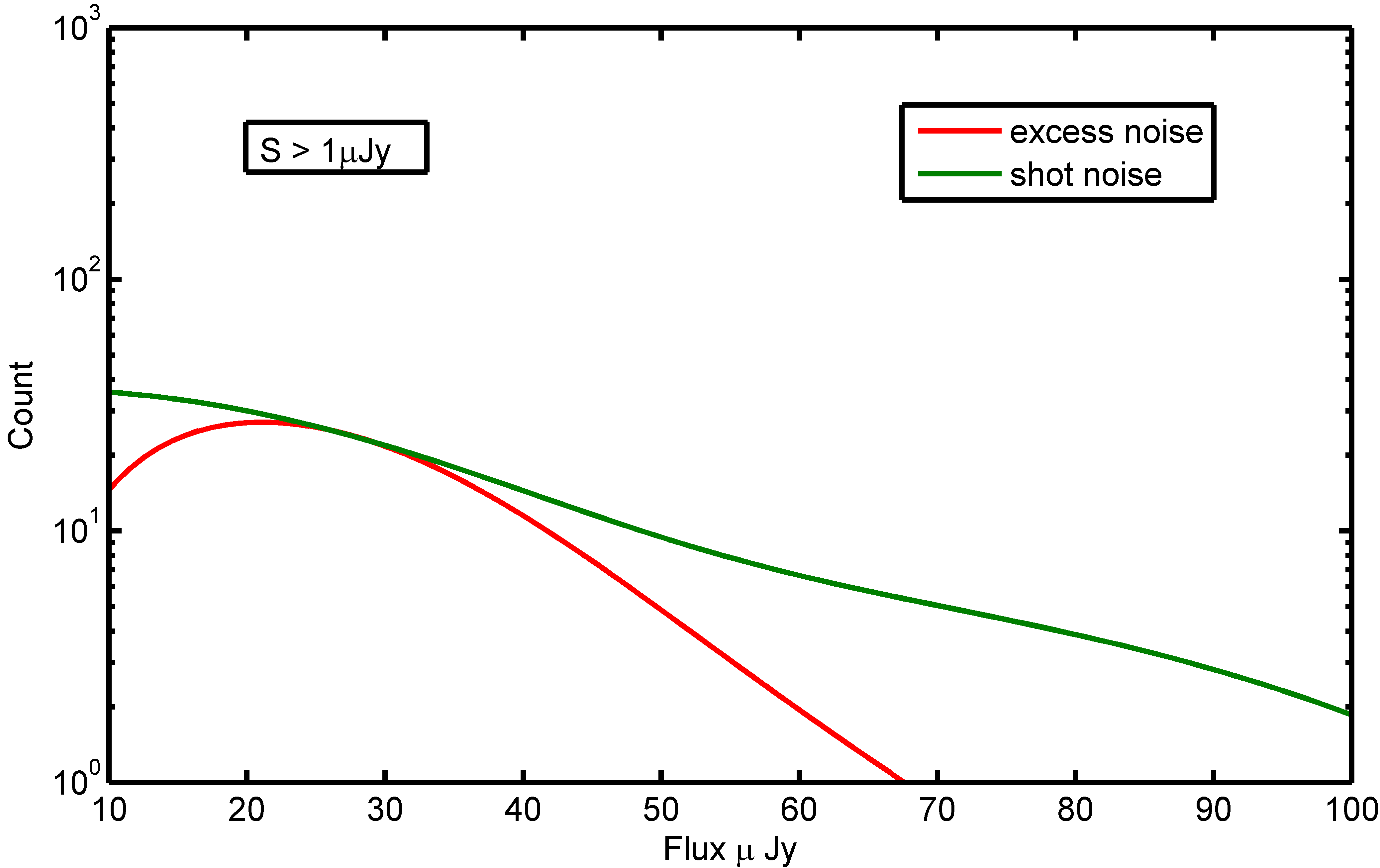}
\caption{The effective noise excess $E_{eff}$ , Gaussian image noise and stacking histogram shot noise as function of flux: Upper plane shows $S_{min}=0.1{\mu}Jy$ case; Lower plane shows $S_{min}=1{\mu}Jy$ case}
\label{fig:signal_win}
\end{figure}

Figure~\ref{fig:signal_win} shows the excess $E_{eff}$ behaviour for
two different minimum flux conditions, $S_{min} = 0.1 \mu$Jy and
$S_{min} = 1 \mu$Jy. 
For the case $S_{min} = 1 \mu{Jy}$, the key point is that $E_{eff}$ has been
overtaken by the shot noise over the FULL flux range. 
This suggests
that we would not have a high enough signal-to-noise ratio to `detect'
the confusion contribution and, as a result, the reconstructed number
counts will not be much affected by confusion. 
This argument supports
the previous findings of Z15, where in the $S_{min} = 0.1 \mu{Jy}$
image the reconstructed
$\frac{{\rm d}^2N_s}{{\rm d}S{\rm d}\Omega} \left(S\right)$ function
is significantly biased, but for $S_{min} = 1 \mu{Jy}$ it is not.

From Eq.\ref{eq:stk_confusion_high}, we see the stacking PDF
$\mathcal{P}_\text{stk}(S)$ contains three components: (i) the
stacking galaxy number count, (ii) the Gaussian image noise and (iii)
the `confusion effect'. If we multiply the number of stacking galaxies
by a factor of $N$, and keep the image noise the same, then the
effective noise excess $E_{eff}$ is amplified by $N$, but the shot
noise is only amplified by $\sqrt{N}$.  For the above case, when we
change $S_{min}$ from $1\mu\mathrm{Jy}$ to $0.1\mu\mathrm{Jy}$, the
number of stacking galaxies is amplified by a factor of 4.

The shot noise $\mathcal{N}_p$ from the stacking pixel-flux histogram
can also be changed by varying the Gaussian image noise or the PSF
resolution. However, the underlying relations are not straightforward
to quantify.  Numerical calculation of the above equations, which will
not be discussed here, would be required.

We also measured the rms $\sigma_o$ of the observed image
pixel-fluxes, and estimated the rms $\sigma_c$ of the noise-free
source confusion\citep{2012ApJ...758...23C} via,

\begin{equation}
 \sigma_c=\sqrt{\sigma_o^2-\sigma_n^2}\;.
\end{equation}

\noindent
The results are shown in Table~\ref{tab:sigma}. Notice that the rms of
the noise-free source confusion is not very sensitive to the minimum
source flux $S_{min}$.  In contradiction to what has previously been
assumed in the community, $\sigma_c$ is a poor estimator of the
eventual bias in the inferred stacking number counts. (We see this
because the bias in the inferred number count changes even when
$\sigma_c$ is roughly constant.)



\begin{table}
\centering
\caption{Confusion noise $\sigma_c$ values for different simulations.
\label{tab:sigma}}
\begin{tabular}{cc}
$S_{\mathrm{min}}$ & $\sigma_c/\mu$Jy \\
\hline
1.0 & 3.718 \\
0.1 & 3.724 \\
0.02 & 3.724 \\
\end{tabular}
\end{table}

The key issue to the stacking method
is whether the effect noise distribution is significant different from Gaussian distribution,
rather than from the thermal map noise beaten down by the noise-free source confusion. 
This difference is compared to the shot noise $\mathcal{N}_p$ from the stacking histogram.
When the difference is overtaken by the shot noise, we can use the ordinary stacking method from Z15, otherwise we can not.

Finally, comparing $E_{eff}(S)$ with $\mathcal{N}_P(S)$, we can describe the more general limitations
of the ordinary stacking method.  We assume that the shape (not the
amplitude) of the stacking-population number counts does not change
much with respect to the SKADS-$S^3$ number counts.  For a
$1-\mathrm{deg}^2$, 6-arcsec-resolution, $16.2$ $\mu$Jy noise image whose galaxies fluxes lie in the range
$0.02\mu\mathrm{Jy}<S<85\mu\mathrm{Jy}$, we found that the upper limit
to the surface density of sources for adopting the ordinary stacking
method is $10,000$ stacking galaxies(i.e. $E_{eff}(S)$ touches $\mathcal{N}_P(S)$).  For the
cases that the density of stacking galaxies is
$>10,000$, the stacking method with confusion must be
used for extracting unbiased counts.

%% file: simulations.tex
\section{Simulation}
\label{sec:sims}

The Square Kilometre Array Design Studies SKA Simulated Skies
(SKADS--S$^3$) simulation \citep{wilman2008,wilman2010} is a
semi-empirical model of the extragalactic radio-continuum sky covering
an area of $400~\mathrm{deg}^2$ , from which Z15 extracted a $1 \mathrm{deg}^2$ catalogue
at $1.4$ GHz for the purposes of testing their method. The simulation,
which is the most recent available, incorporates both large- and
small-scale clustering and has a flux limit of about $0.01\,\mu$Jy. We
undertook several tests as follows.

For the present study, we adopt the same VLA-like mock survey strategy
as Z15, i.e. a 1.4-GHz, 1-deg$^2$ survey with a gaussian noise with
$\sigma=16.2\,\mu$Jy and a 6-arcsec gaussian FWHM
synthesized beam/PSF. This setup follows the VLA-VIRMOS observations
by \cite{bondi2003}. 

In total $374,061$ sources have been injected into
the simulated image taken from SKADS, whose minimum injected
source flux is $0.02\,\mu$Jy\footnote{The simulation limit is about  $0.01\,\mu$Jy, and we choose $0.02\,\mu$Jy to be a conservative cutoff}. 
We further set a source flux cutoff at $85\,\mu$Jy, while we assume the sources with flux larger than this value can be masked out.
The flux cutoff setting will not affect our finally result.

To match with ancillary optical/infra-red data (e.g.~the VIDEO survey;
\citealt{jarvis2013}), we select the star forming galaxies whose K-magnitude $< 24$ in the simulated catalogue.
In total there are $149,516$ stacking sources left under this selection(hereafter 'selection-H').

However, $149,516$ stacking galaxies out of $374,061$ are way too large to use the low density stacking method.
For this reason, we create another stacking selection by adopting a further constraint on redshift $0.5<z<1.0$.
After the redshift selection, only $18,110$ galaxies are left(hereafter 'selection-L'). 
This selection fulfills the requirements of the low density stacking method.

It is also important to decide the size of image pixel. In order to
simplify the analysis, we expect the bins of the image pixel fluxes histogram to
be independent from each other, i.e. we want to remove the pixel-to-pixel
correlation\footnote{Ideally, the pixel-to-pixel correlation will not change the best-fit stacking galaxy number count, but extends the uncertainty of the number count.}.  This requires that the pixel size is comparable to the PSF
resolution \citep{vernstrom2014}. However, we also want to reduce the offset of the stacking galaxy from the center of its own pixel.  
In our simulation, we choose the pixel size to be $1$ arcsec.

\subsection{Modelling differential number count }
\label{sec:systematics:dnds}

There are many ways of modelling the differential number-count
function ${\rm d}^2N/({\rm d}S{\rm d}\Omega)$.  From the mathematical point of view,
the most straightforward model is the polynomial in log-log space
 (see e.g.~\citealt{optimal-binning,vernstrom2014}). However, the
${\rm d}^2N/({\rm d}S{\rm d}\Omega)$ function is over-sensitive to
the higher-order parameters in the polynomial model, which need a lot
of attention in the sampling process, as well as the prior range.

The pole/node-based model (see e.g.~\citealt{vernstrom2014}) fixes the
position in $\log_{10}(S)$ of a fixed number of nodes, and we fit for
the node amplitudes. Between the nodes the count is interpolated in
log space to ensure a continuous function. In this method, the node
amplitudes depend not only on the underlying source count but also on
the number, or spacing, of the nodes, and also the type of
interpolation used between the nodes.  Therefore, given a differential
number count, the choice of node number and positions has considerable
impact on the fit results.

Similar to the pole/node-based model, the multi-power-law model is
also based on a small number of break points connected by power law
segments.  Unlike the node-based model, the flux positions of the
break points are not fixed, and the physical meaning of each fitting
parameter is obvious, but the parameters of the model are highly
correlated and high signal-to-noise features tend to attract the free
break positions.

Overall, we adopt the multi-power-law model in the following analysis
for simplicity.  We choose two cases: one-break power-law model and two-breaks power-law model(Models B and C from
Z15). Model A in Z15 does not have enough features to characterize the
shape of the SKADS differential source count, and model D has too many
correlated parameters. The one-break power-law model is defined as

\begin{multline}
\label{eqn:tpl}
\frac{{\rm d}^2N}{{\rm d}S{\rm d}\Omega}
\left(C,\alpha,\beta,S_0,S_{min},S_{max}\right)\\
=
\begin{cases}
CS^{\alpha} & S_{min}<S<S_0\\
C\,S_0^{\alpha-\beta\,}S^{\beta} & S_0<S<S_{max}\\
    0      & \mathrm{otherwise}\\
\end{cases},
\end{multline}

\noindent with a parameter vector
$\mathbf{\Theta_B}=\{C,\alpha,\beta,S_0,S_{\mathrm{min}},S_{\mathrm{max}}\}$.
The two-breaks power-law model incorporates another break in the power law, so
$\mathbf{\Theta_C}=\{C,\alpha,\beta,\gamma,S_0,S_1,S_{\mathrm{min}},S_{\mathrm{max}}\}$. Priors
on the different model parameters are discussed in section
\ref{sec:bayes:priors}.

\subsection{Priors}
\label{sec:bayes:priors}

Our conservative priors are listed in Table~\ref{table:priors}. Note
that we have assumed equiprobable models \textit{a priori},
i.e.~$\mathcal{Z}_1\left(\mathrm{\mathbf{D}}|H_1\right)=\mathcal{Z}_2\left(\mathrm{\mathbf{D}}|
  H_2\right)$.

\begin{table}
\centering
\caption{Priors $\mathit{\Pi}\left(\mathbf{\Theta}| H\right)$ adopted
  here.\label{table:priors}}
\begin{tabular}{ll}
\hline
Parameter \T\B & Prior  \\
\hline
$C/\mathrm{sr^{-1}Jy^{-1}}$ \T\B & log-uniform $\in \left[10^{-5},10^9\right]$ \
\\
$\alpha_j$ \B & uniform $\in \left[-2.5,-0.1\right]$ \\
$S_{\mathrm{min}}/\mu$Jy \B & uniform $\in \left[0.001,1.5\right]$ \\
$S_{\mathrm{max}}/\mu$Jy \B & uniform $\in \left[1.5,5.0\right]$ \\
$S_{0,1}$ \B & uniform $\in \left[S_{\mathrm{min}},S_{\mathrm{max}}\right]$ \\
$S_{0,1}$ \B & further require $S_0<S_1$ \\
$\sigma$ \B & uniform $\in \left[0.5,2.0\right]\sigma_{\mathrm{survey}}$ \\
$\mathcal{Z}_i$ \B & equiprobable \\
\hline
\end{tabular}
\end{table}

%% file: results.tex
\section{Results}
\label{sec:results}

We now test the methods developed in the previous sections.  We set
about extracting binned source counts at the positions of the selected
sources, as described in Section \ref{sec:sims}, for the two
selections (i.e.~Selection-L and Selection-H) from the same image.
Additionally, in order to estimate the contribution to the inferred
differential number count from confusion, we also generated a
histogram of all image pixel fluxes as the input for a $P(D)$
modeling of the whole image.

It is worth pointing out that the image noise is relatively
independent of the parameters of the number-count model, and is very
well constrained by the pixel-flux histogram. 
In order to speed up computation, therefore, we have not fitted
simultaneously for the image noise, but our software does already have
this capability (as in Z15).

\begin{figure*}
	\includegraphics[width=\textwidth]{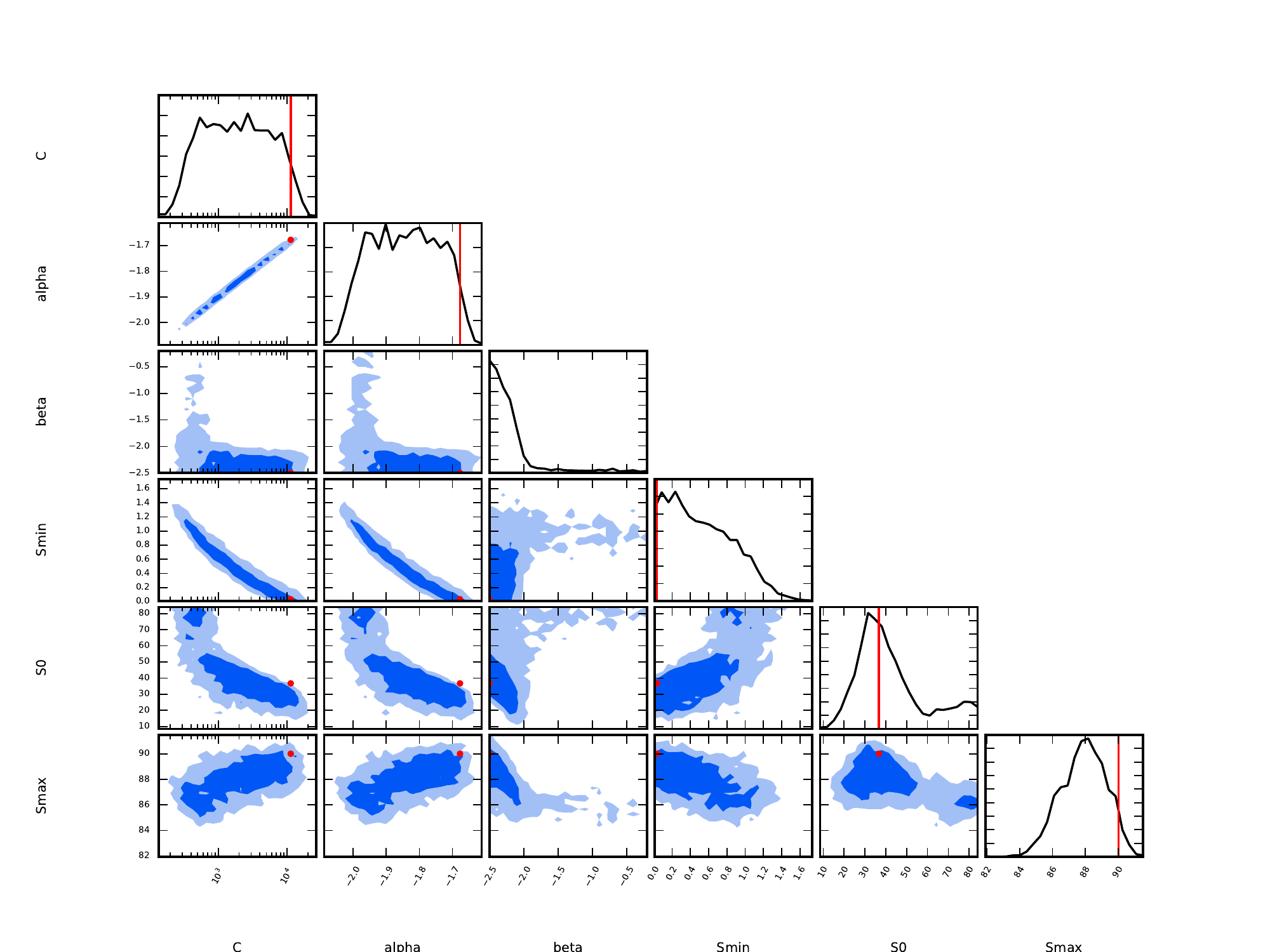}
        \caption{Posterior probability distribution for a $P(D)$ fit
          to the differential number count of the non-stacking sources
          using a single-break power-law model. The red dots shows the
          maximum-likelihood parameters and the 68 and 95~per~cent
          confidence limits are respectively indicated by the dark and
          light shaded regions.}
      \label{fig:res_pofd_1bpl}
\end{figure*}

The one-break power-law posterior probability distribution for the
total galaxy differential number counts via the $P(D)$ analysis is
shown in Fig.~\ref{fig:res_pofd_1bpl}. We clearly see that the break
flux $S_0$ has two preferred values, centering at $30{\mu}$Jy and
$80{\mu}$Jy, which are the two breaks in the differential number
counts of the SKADS-S$^3$ simulation. The relative evidence,
$\Delta\log_e\mathcal{Z}=0.63 \pm 0.27$ indicates that this model is
preferred to a two-breaks power-law model at the $\approx 3$-$\sigma$ level.

\begin{figure}
	\includegraphics[width=0.5\textwidth]{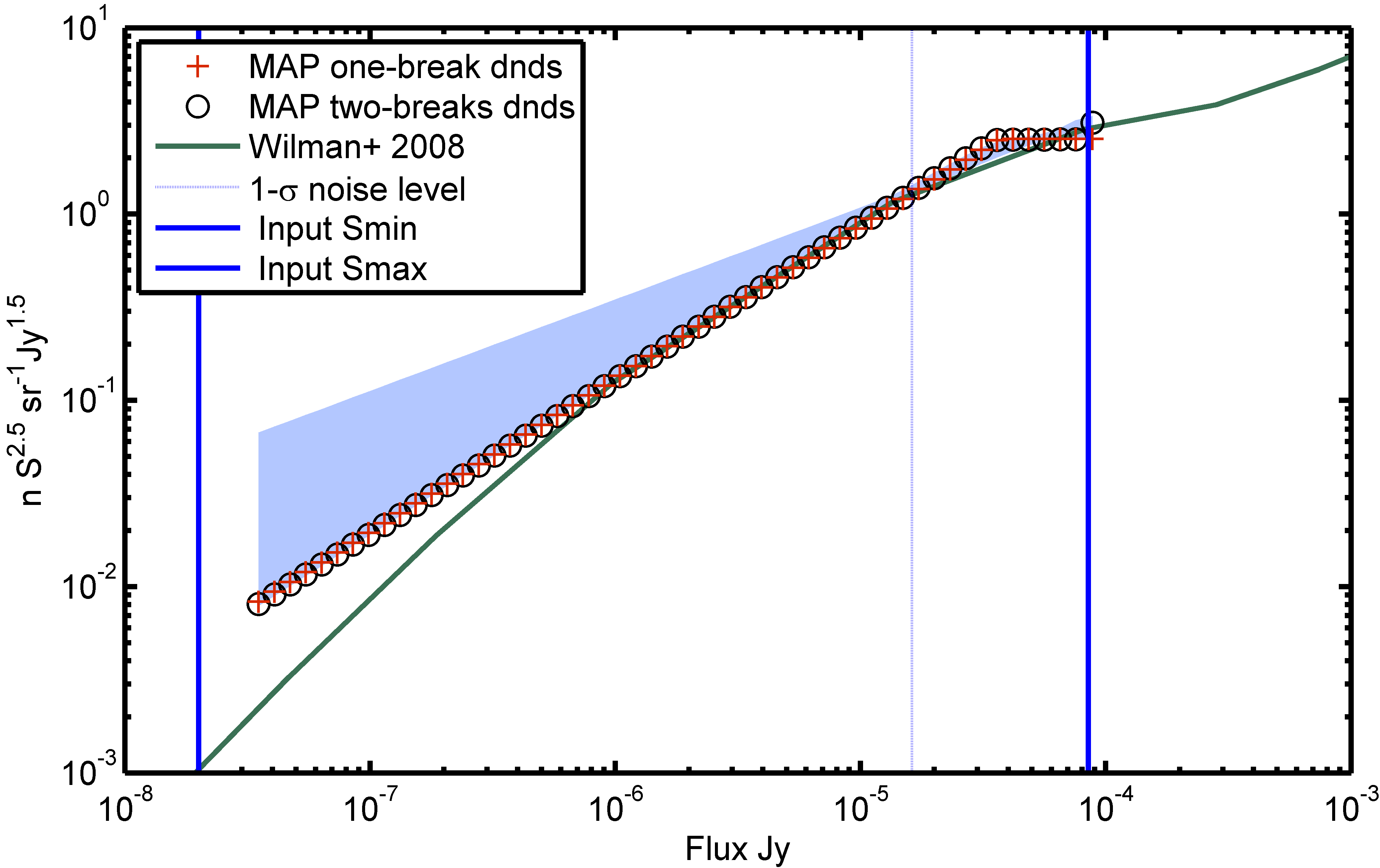}
        \caption{Reconstructed differential number counts for all mock
          galaxies using one-break power-law model(best-fit: red crosses).
          The blue area represents the 95~per~cent
          confidence interval.            
          The black circles is the maximum-likelihood two-breaks power-law differential number counts for comparison.
          The green solid line indicates the mock number counts.
          The blue dot line shows the noise rms, and the blue solid lines shows the flux minimum and maximum of simulation sources.}
      \label{fig:recon_pofd_1bpl}
\end{figure}

Using the above posterior probability distribution, we reconstructed
the total galaxy differential number counts
(Fig.~\ref{fig:recon_pofd_1bpl}). These follow the mock number
counts tightly down to $1{\mu}$Jy. 
It is looks like the number counts at faint fluxes $>1{\mu}$Jy can be fitted better by allowing more breaks in the number counts model. 
However, the fitting results from the two-breaks power-law model show that the break are attracted to high signal-to-noise region which is around the bright end(see Fig.~\ref{fig:recon_pofd_1bpl}). 
A larger image with finer flux resolution and lower noise may help to improve the
fitting at fluxes $S<1{\mu}$Jy by increasing the signal-to-noise ratio in this flux range.

Summarizing a posterior probability distribution is challenging but a
necessary aspect of our two-step approach (see also
section~\ref{sec:simul-fit} below). Ideally, one would
importance-sample from the $P(D)$ posterior distribution so as to
propagate its morphology (i.e.~uncertainties, correlations,
degeneracies, multimodalities, skirts, wings and any other
non-gaussianities). Nonetheless, for simplicity, rather than
considering each parameter on a case-by-case basis, we do try to
describe the posterior distribution using a single statistic, noting
especially that this will not propagate the uncertainties from the
$P(D)$ part of the analysis. We adopt the maximum-likelihood parameter estimates, 
which fit the mock number counts better than the median parameter, 
and for reference these values are given in Table~\ref{tab:pd_fit}.


\begin{table}
\centering
\caption{For the $P(D)$ analysis, maximum \textit{a posteriori}
  parameter estimates for the winning one-break power-law
  model. We have not included parameter uncertainties is this table
  in order to emphasize the fact that these are not propagated to the
  joint $P(D)$ plus stacking step.\label{tab:pd_fit}}
\begin{tabular}{cc}
Parameter & Value \\
\hline
$\hat{C}/\mathrm{sr}^{-1}\mathrm{Jy}^{-1}$ & $1.1\times 10^4$\\
$\hat{\alpha}$ & $-0.17$ \\
$\hat{\beta}$ & $-2.50$ \\
$\hat{S}_{0}/\mu \mathrm{Jy}$ & 36.7 \\
$\hat{S}_{min}/\mu \mathrm{Jy}$ & 0.034 \\
$\hat{S}_{max}/\mu \mathrm{Jy}$ &  90.0 \\
\end{tabular}
\end{table}

\subsection{Stacking analysis (Selection-L)}

For stacking Selection-L, we assume the above reconstructed
number-count function from the full image to be a good approximation
to the non-stacking number counts. Note that the non-stacking galaxies
represent about 95~per~cent of the total number of galaxies.

\begin{figure*}
	\includegraphics[width=\textwidth]{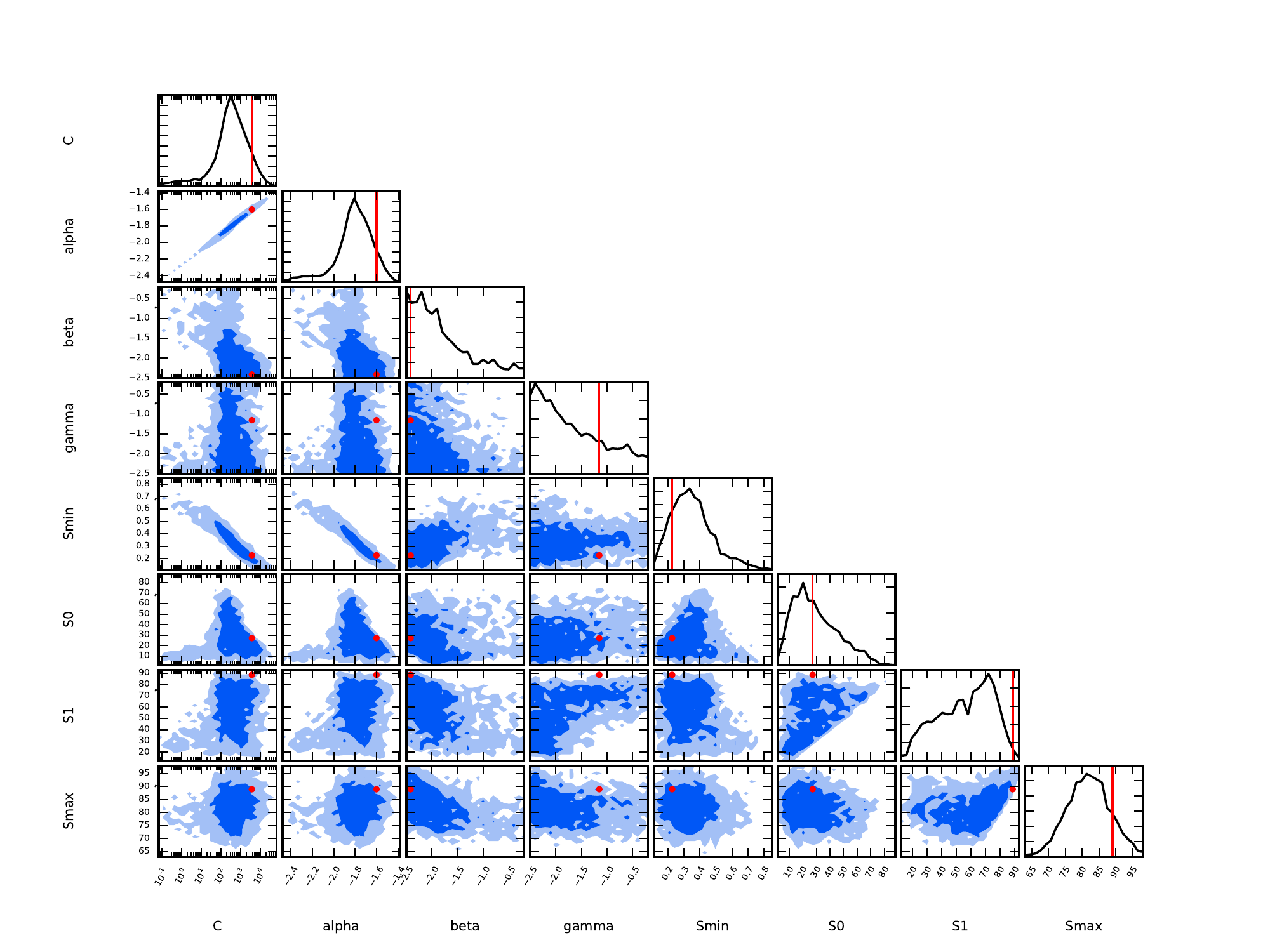}
        \caption{Posterior probability distribution for a low-density
          stacking fit to the differential number count of the
          stacking population using the two-break power-law model. The
          red dots shows the maximum-likelihood parameters and the 68
          and 95~per~cent confidence limits are respectively indicated
          by the dark and light shaded regions.}
      \label{fig:res_stk_2bpl}
\end{figure*}

Assuming the parameters from Table~\ref{tab:pd_fit} as the
prescription for the number counts of the non-stacking galaxies, we
implemented the low-density stacking method (Section
\ref{sec:algorithm}) in order to extract the number counts of the
stacked population. We compared the fits of the one-break power-law
stacking model and the two-breaks power-law stacking model via the
Bayesian evidence (Table~\ref{tab:evidence_sel_L}). The model
evidences are fully consistent with each other within
uncertainties; we choose to adopt the two-breaks power law as the
marginal winner since strictly its evidence is the higher. The
posterior probability distribution for two-breaks power-law stacking
model is shown in Fig.~\ref{fig:res_stk_2bpl}.

\begin{table}
\centering
\caption{Selection-L (low-density method): Nested Sampling Global $\log_e$-evidence for the different
  power-law models.
\label{tab:evidence_sel_L}}
\begin{tabular}{ccc}
 Breaks & Parameters & $\log_e$-evidence \\
\hline
1 & 6 & $-118.0604\pm0.1740$ \\
2 & 8 & $-118.0598\pm0.1745$ \\
\end{tabular}
\end{table}

\begin{figure}
	\includegraphics[width=0.5\textwidth]{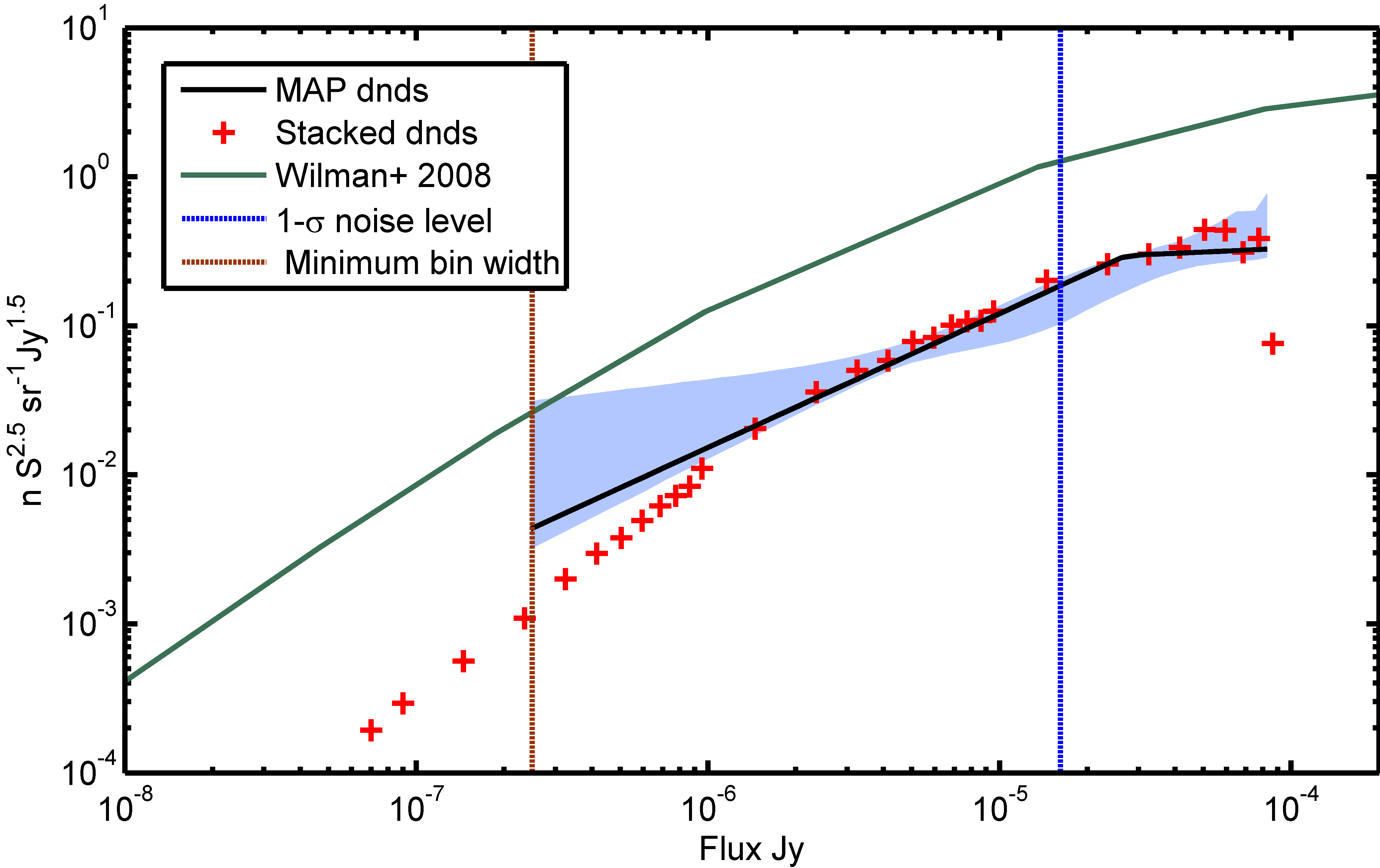}
        \caption{Reconstructed differential number counts for the
          Selection-L population using the low-density method with
          two-breaks power-law model(best-fit: solid blue line).
          The blue area represents the 95~per~cent
          confidence interval. 
          The red crosses indicate the mock 
          Selection-L stacking-population number counts. 
          The blue dot line shows the noise rms, and brown dot line shows the minimum stacking pixel fluxes histogram bin width(resolution).
          }
      \label{fig:recon_stk_2bpl}
\end{figure}

From this we reconstructed the number counts of the stacked
population (Fig.~\ref{fig:recon_stk_2bpl}). The reconstructed counts
are fully consistent with the mock counts of the selection-L
sources within the 95~per~cent confidence level, except for a slight
offset at the very faint end ($S<1\mu$Jy).

\begin{figure}
	\includegraphics[width=\linewidth]{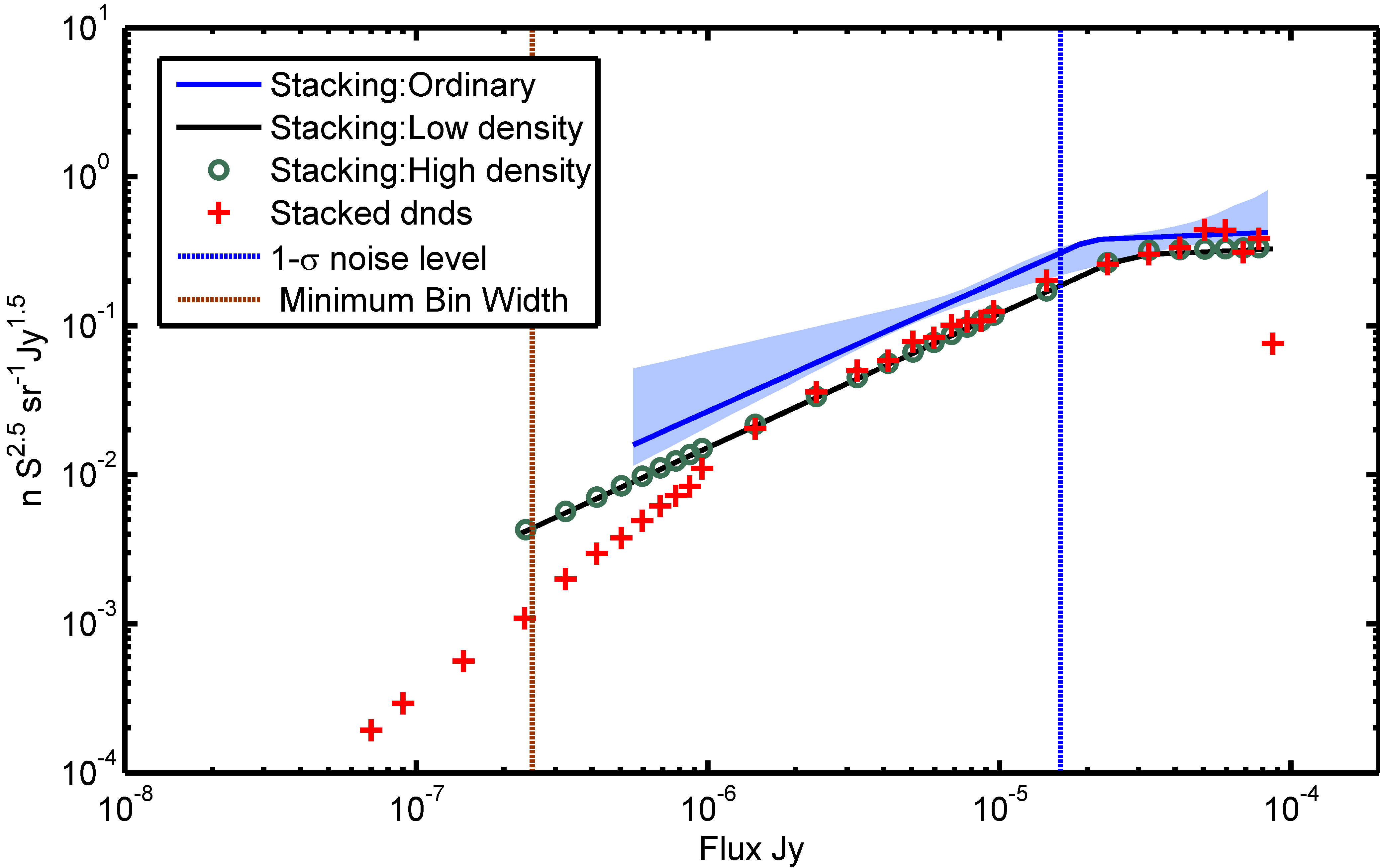}
        \caption{Reconstructed stacking differential number counts
          compared to the ordinary stacking method that ignores
          confusion (best-fit: solid blue line).
          The blue area represents the 95~per~cent
          confidence interval. The red crosses indicate the mock 
          Selection-L stacking-population number counts, with the two-breaks
          power-law reconstructions using the low-density method
          (section \ref{sec:algorithm}) are shown with a black solid
          line and the high-density method with green circles.}
      \label{fig:oldstk}
\end{figure}

Fig.~\ref{fig:oldstk} shows a comparison of the different available
stacking methods. The ordinary method gives a significantly-biased
reconstruction (in agreement with the findings of Z15). The
low-density and high-density methods give almost identical results,
and fit the stacked number counts reasonably well.

\subsection{Stacking analysis (selection-H)}
\label{sec:results_high}

\begin{figure*}
	\includegraphics[width=\textwidth]{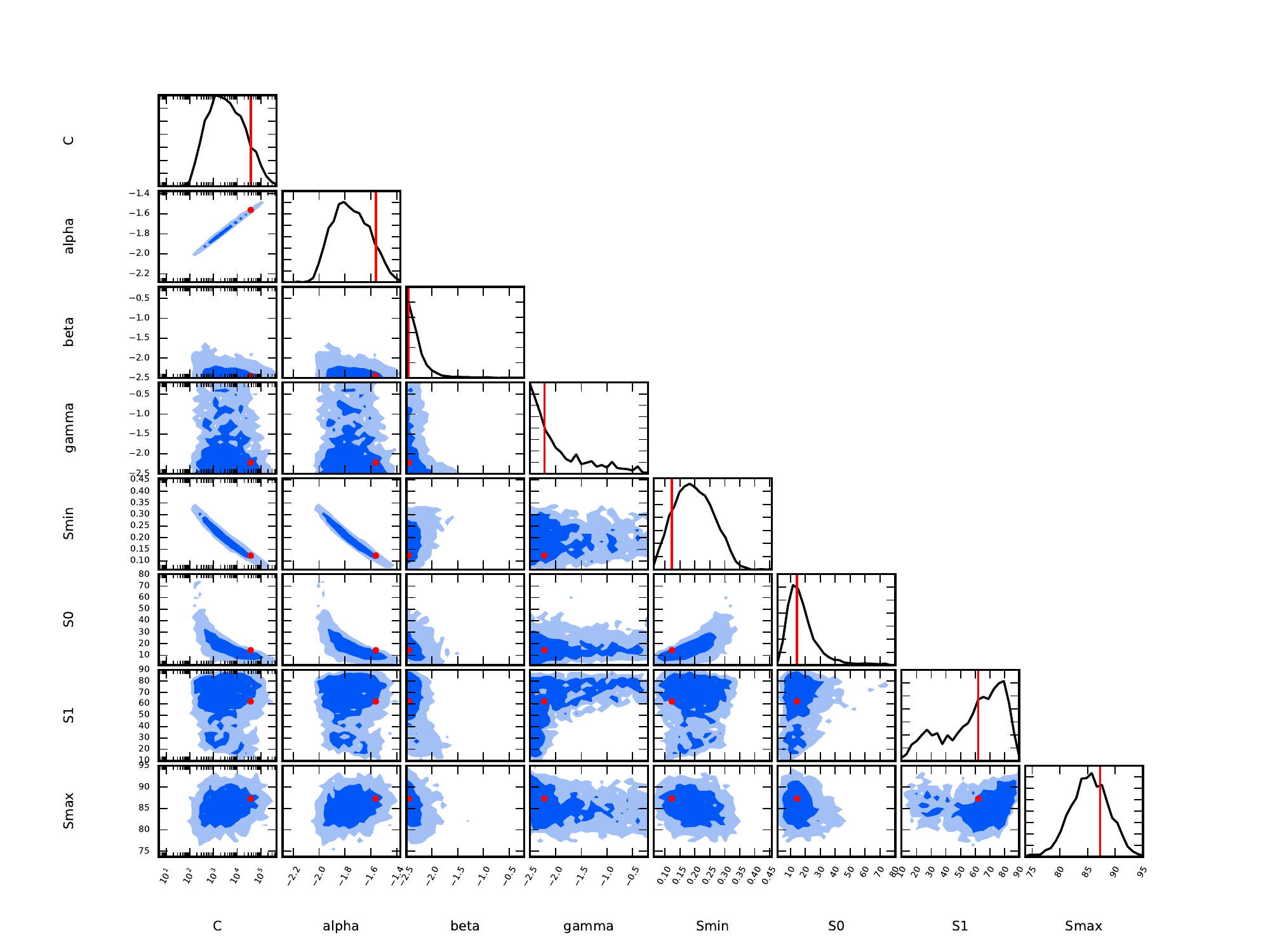}
        \caption{
          Posterior probability distribution for a high-density
          stacking fit to the differential number count of the
          stacking population using the two-breaks power-law model. The
          red dots shows the maximum-likelihood parameters and the 68
          and 95~per~cent confidence limits are respectively indicated
          by the dark and light shaded regions.
        }
      \label{fig:old_map_res_stk_2bpl}
\end{figure*}

For stacking selection-H, non-stacking galaxies represent about
60~per~cent of the total population, so that the number counts of the
total population is no longer a good approximation to those of the
non-stacking galaxies. Under these circumstances the low-density
method is unsuitable, and could yield biased reconstructed number
counts. Hence we use the high density method,
Table~\ref{tab:evidence_sel_H} giving the relative evidences for
different power-law source-count models fitted to the data. The data
prefer the two-breaks model, whose posterior probability distribution
is shown in Fig.~\ref{fig:old_map_res_stk_2bpl}.


\begin{table}
\centering
\caption{Selection-H (high-density method): Nested Sampling Global $\log_e$-evidence for the different
  power-law models.
\label{tab:evidence_sel_H}}
\begin{tabular}{ccc}
 Breaks & Parameters & $\log_e$-evidence \\
\hline
 1 & 6 & $-123.8 \pm 0.2$ \\
 2 & 8 & $-122.9\pm0.2$ \\
\end{tabular}
\end{table}

\begin{figure}
	\includegraphics[width=\linewidth]{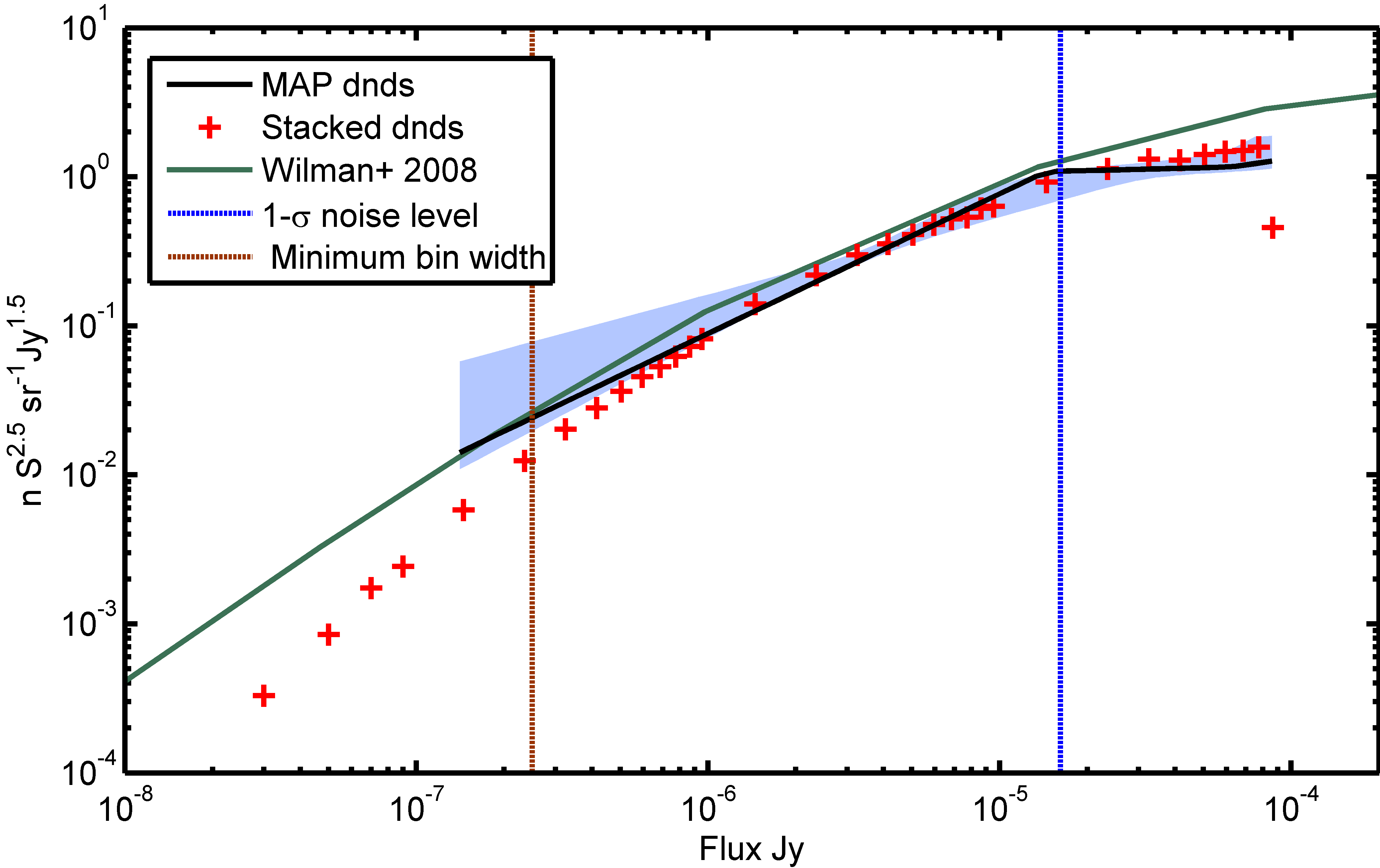}
        \caption{
          Reconstructed differential number counts for the
          Selection-H population using the high-density method with
          two-breaks power-law model (best-fit: solid blue line).
          The blue area represents the 95~per~cent
          confidence interval. 
          The red crosses indicate the mock 
          Selection-H stacking-population number counts. 
          The blue dot line shows the noise rms, and brown dot line shows the minimum stacking pixel fluxes histogram bin width(resolution).
          }
      \label{fig:old_map_recon_stk_2bpl}
\end{figure}

Reconstructing the source counts of the stacked population from the
posterior probability distribution(Fig.~\ref{fig:old_map_recon_stk_2bpl}), at the 95-per-cent level
these are fully consistent with the mock counts for the
selection-H sources, except for a slight offset at the very faint end,
$S<1\mu$Jy.

\begin{figure}
	\includegraphics[width=\linewidth]{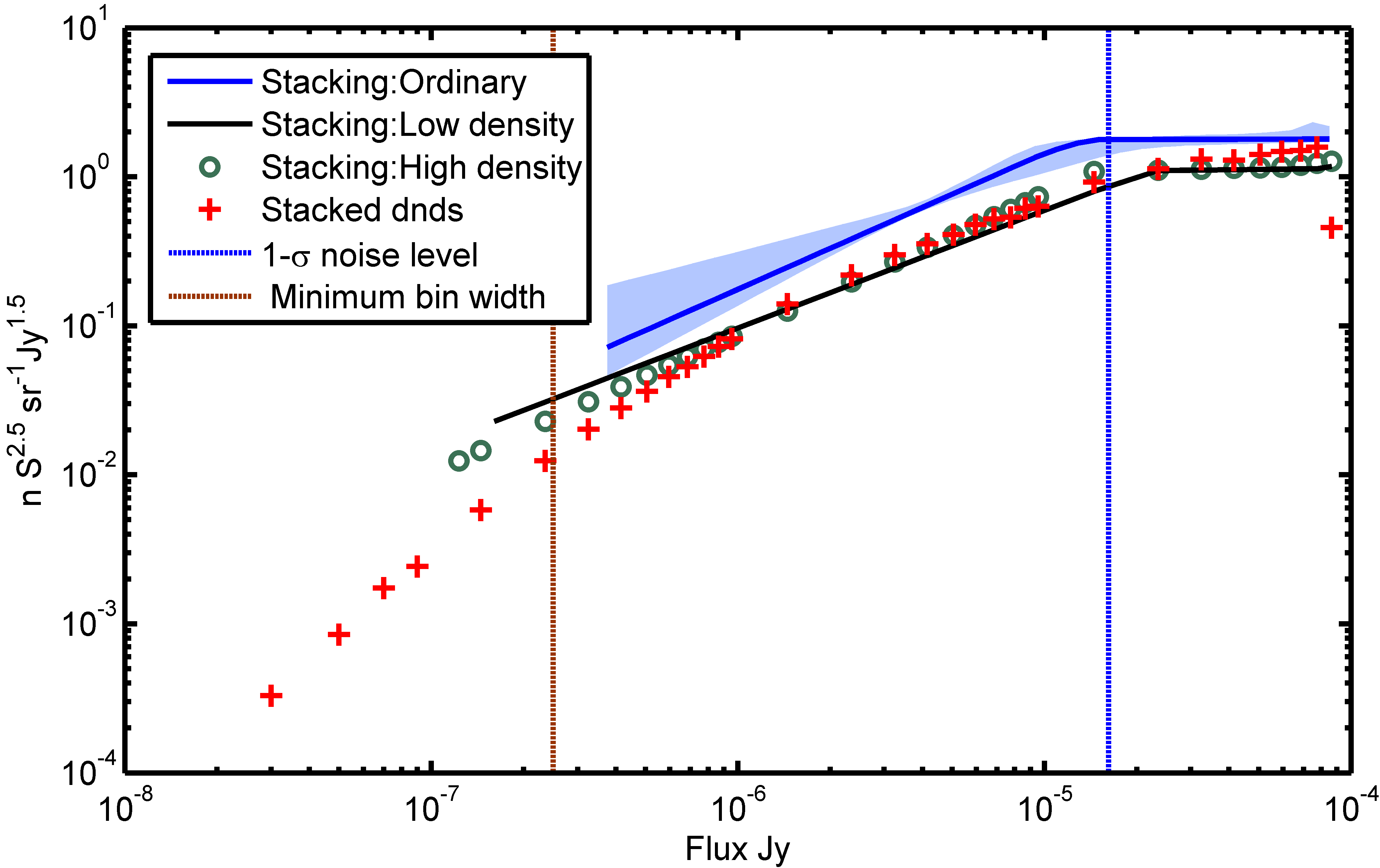}
        \caption{
          Reconstructed selection-H stacking differential number counts 
          compared to the ordinary stacking method that ignores
          confusion (best-fit: solid blue line).
          The blue area represents the 95~per~cent
          confidence interval.
          The red crosses show the mock
          stacking-population number counts, with the two-breaks
          power-law reconstructions using the low-density method
          (section \ref{sec:algorithm}) are shown with a black solid
          line and the high-density method with green circles.
        }
      \label{fig:old_map_compare}
\end{figure}

In Fig.~\ref{fig:old_map_compare}, the reconstructed
differential number counts using the ordinary method (solid blue line)
is biased again, for the same reason of Z15. 
Besides, the low density method(black line) is below the mock number counts in the range $S>1\mu$Jy. 
This is probably due to the fact that
we have overestimated the non-stacking source count using an
inappropriate approximation.

As a check, we calculated the difference between the mock number counts
and the different reconstructed number counts, via $\chi^2$.
We define $\chi^2$ as
\begin{equation}
 \chi^2\equiv \frac{1}{N}\sum_{i=1}^N (\frac{ \text{reconstructed dnds}-\text{mock dnds} }{\frac{1}{2}\times 95\% CI})_i^2 \;,
\end{equation}
where, $N$ is the total number of the number counts bins.

The $\chi^2$ values in Table \ref{tab:chi2} indicate the high density method gives the
closest reconstructed number counts.

\begin{table}
\centering
\caption{$\chi^2$ values for the reconstructed different number counts with
  respect to the mock number counts(for 28 bins with $S>1\mu$Jy). 
  \label{tab:chi2}}
\begin{tabular}{c|c|c|c|}
Method & $\chi^2$  \\
\hline
Ordinary (Z15) & $43.32$ \\
Low-density  & $3.44$ \\
High-density  & $1.92$ \\
\end{tabular}
\end{table}

\section{Simultaneous fitting total source count and stacking source count}
\label{sec:simul-fit}

In general, a simultaneous analysis via $P(D)$ \textit{and} stacking is preferred
to our adopted two-step approach (see section~\ref{sec:recipe}) in
which a fitted $P(D)$ model is used as the input to a joint $P(D)$
plus stacking analysis. 
However, this method is time consuming, and requires a large number of simulations
 to understand the correlations between the total pixel-flux histogram 
and stacking pixel-flux histogram.

In order to understand the interplay between the parameters of the
total galaxy differential number-count model and the
stacking-population differential number-count model, we fitted two
differential number-count models simultaneously with a simplified joint
likelihood. This was simply the product of the two likelihoods:

\begin{equation}
\mathcal{L}_\text{joint}=\left(\prod_i \frac{I_i^{k_i}}{k_i !}\exp(-I_i)\right)\left(\prod_j \frac{J_j^{k_j}}{k_j !}\exp(-J_j)\right).
\end{equation}

\begin{figure}
	\includegraphics[width=\linewidth]{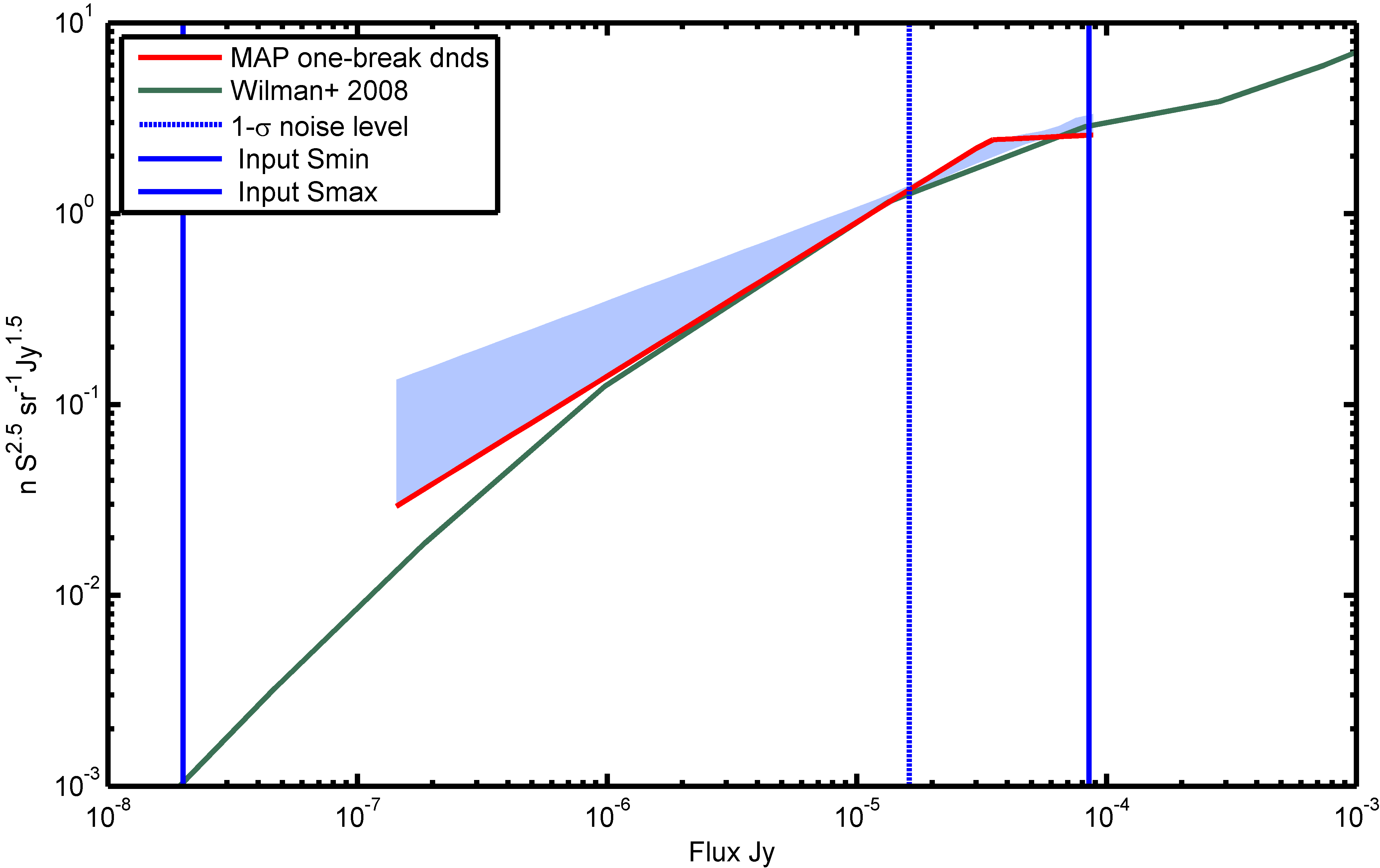}
        \caption{
          Reconstructed differential number counts for the
          total galaxies using the joint method with
          single-break power-law model (best-fit: solid red line).
          The blue area represents the 95~per~cent
          confidence interval. 
          The green solid line indicate the mock number counts.
          \newline
          \newline
          \newline
          }
      \label{fig:highstk-joint_pd}
\end{figure}

\begin{figure}
	\includegraphics[width=\linewidth]{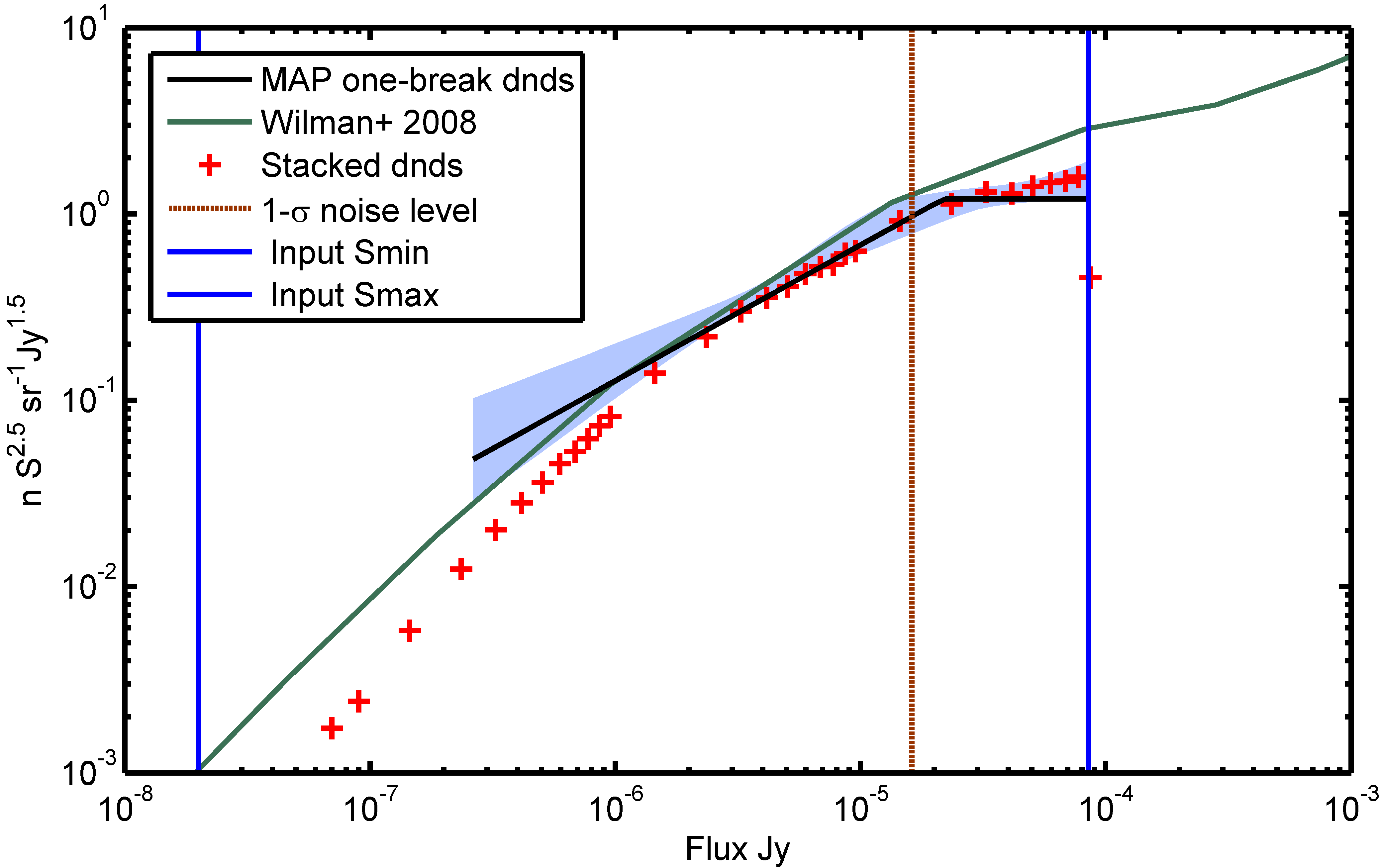}
        \caption{
          Reconstructed differential number counts for the
          selection-H stacking population using the joint method with
          one-break power-law model(best-fit: black solid line).
          The blue area represents the 95~per~cent
          confidence interval. 
          The red crosses indicate the mock stacking-population number counts.
        }
      \label{fig:highstk-joint_stk}
\end{figure}

\noindent
This simplified joint likelihood assumes independence between the total pixel-flux histogram 
and stacking pixel-flux histogram.  
Under this assumption, we tested the joint method with our earlier
simulation.  The fitting results of are shown in
Fig.\ref{fig:highstk-joint_pd} and Fig.\ref{fig:highstk-joint_stk} for
the reconstructed full-image source count and the stacking-population
count respectively. The joint fitting successfully recovers both the
total-galaxy differential number counts and the stacking-population
differential number counts.


\begin{figure*}
	\includegraphics[width=\linewidth]{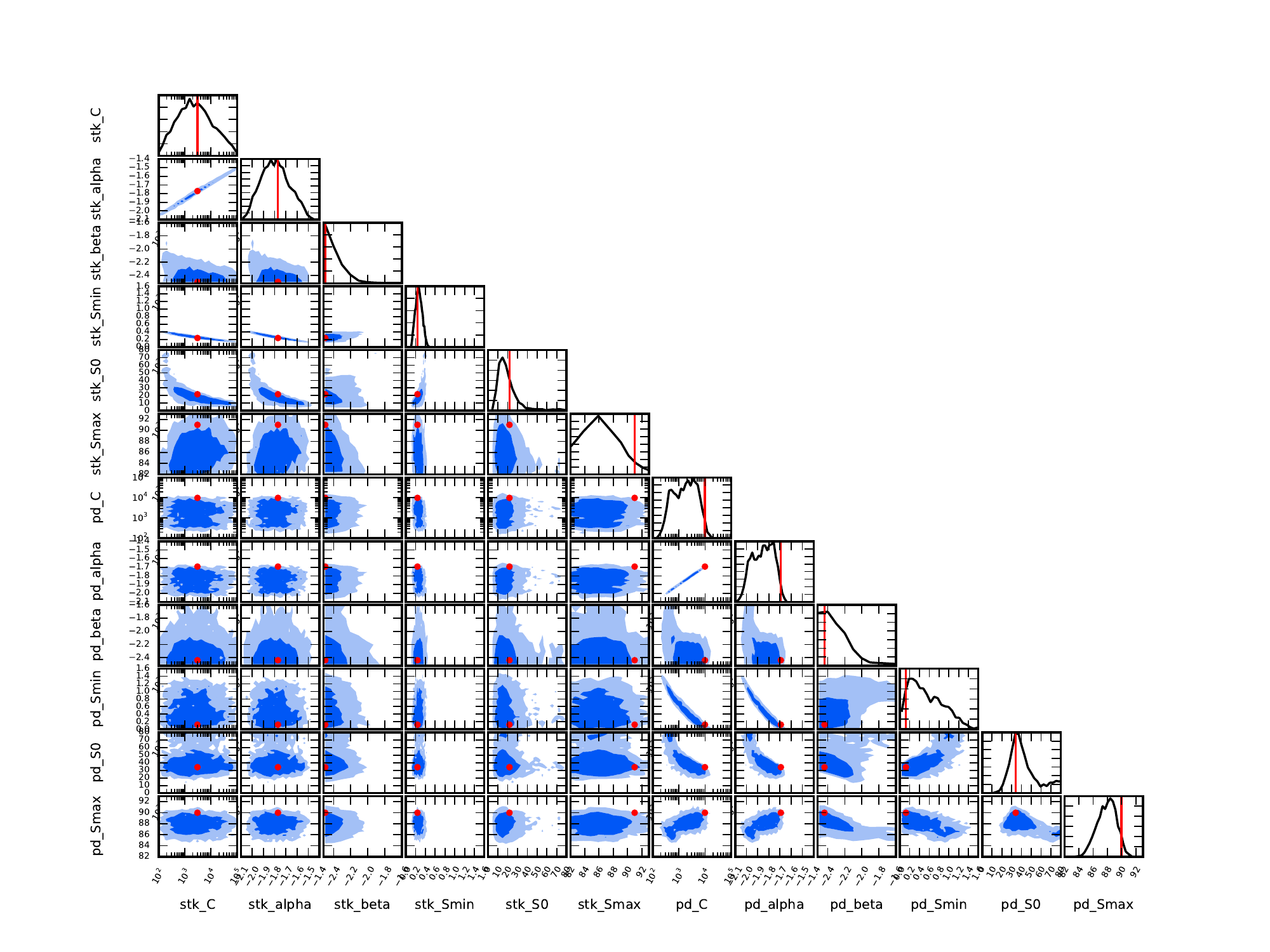}
        \caption{
          Posterior probability distribution for a joint 
          fit to the differential number counts of the
          stacking population and total galaxies.
          we used the one-break power-law models for the two number counts. 
          The red dots shows the maximum-likelihood parameters and the 68
          and 95~per~cent confidence limits are respectively indicated
          by the dark and light shaded regions.
        }
      \label{fig:triangle-highstk-joint}
\end{figure*}

From the posterior probability distribution
(Fig.\ref{fig:triangle-highstk-joint}), we see that significant
correlations solely exist \textit{internally} to the parameters of the
two number-count models. The correlations between the number counts
models are less significant.
The $P(D)$ analysis better constrains $C$, $\alpha$ and
$S_{\mathrm{max}}$ than does stacking. $C$ and $\alpha$ are strongly
correlated parameters and together they fix the general shape of the
number count.  Since there are more pixels for the $P(D)$ analysis
than for the stacking analysis, a stronger constraint on the total
galaxy number counts is expected. However, not all of the $P(D)$
parameters are better constrained.  The stacking analysis gives
tighter constraints on $\beta$, $S_{\mathrm{min}}$ and $S_0$, which
may be because the actual number of `breaks' for stacking-population
model is less than for the total-galaxy model.  In another words, the
actual stacking-population number counts are closer to a single-break
power-law model.

%% file: discussion.tex
\section{Discussion}
\label{sec:discussion}

The reconstructed number counts $S_{min}$ in Fig.\ref{fig:recon_stk_2bpl} and
Fig.\ref{fig:old_map_recon_stk_2bpl} are far from the $S_{min}$ of the stacked number counts.
The posterior probability distribution figures show a strong correlation between the faint slope $\alpha$ and the minimum flux $S_{min}$.
The faint slop parameter $\alpha$ is driven by the slop at brighter fluxes(high signal-to-noise ratio).
As a result, the $S_{min}$ parameter is over estimated. 
Adding one extra break to the number counts model does not solve the problem.
Because the extra break is always attracted to the high signal-to-noise flux region.
To improve this situation, we need to increase the signal-to-noise ratio at faint flux, such as reducing the instrumental noise.

It is also worth mentioning that the statistical uncertainty in the
reconstructed stacking number counts is underestimated whenever the
stacking-pixel fluxes are \textit{not} statistically independent.
This happens when the surface density of stacking galaxies is so high
that the stacking-pixel values are correlated via the beam (or
PSF). Selection-H has 149,516 sources, and the synthesized-beam
resolution is 6\,arcsec.  On average, each source maintains a
$5.3$-arcsec radius circle.  This radius is 2--3$\sigma$ of the PSF,
which corresponds to a weight of 1~per~cent in Gaussian PSF.  
In principle, the uncertainty in the
selection-H stacking results are underestimated because the
actual number of independent pixels must be fewer than 149,516, but this
effect is negligible.

%% file: conclusions.tex
\section{Conclusions}
\label{sec:conclusions}


\begin{enumerate}

\item We have extended the `ordinary' Bayesian stacking technique to
  include the full effects of source confusion, an enhancement that is
  highly applicable to data from forthcoming confusion-limited
  radio-continuum surveys such as MIGHTEE.

\item We have derived two core probability density functions
  (Eq.\ref{eq:stk_confusion_ind} and Eq.\ref{eq:stk_confusion_high})
  that describe the stacking analysis including confusion. One PDF
  uses the non-stacking galaxy number counts, and the other uses the
  number counts for the total galaxy population.

\item We applied these two new stacking methods to synthesized images
  based on the SKADS simulation. With the new methods, the
  reconstructed number counts are fully consistent with the injected
  number counts, while number counts reconstructed via the ordinary
  stacking method had been biased at the 95-per-cent confidence level.

\item While it had previously been assumed in the literature that the
  confusion contribution to a stacking experiment was non-negligible
  if the `confusion noise' was much less than the map thermal
  noise. The conventional confusion noise rms ($\sigma_c$) was shown
  to be a poor estimator for quantifying the impact of confusion on
  stacking analyses. The key issue is whether the total noise is
  significantly different from Gaussian. We provide a new heuristic
  thaf fulfils this role.

\item A joint analysis of $P(D)$ and stacking, assuming independence,
  has allowed us to study the interplay between the parameters of the
  total galaxy number-count and the stacking-population number-count.
  We found significant correlations solely exist \textit{internally}
  to the parameters of the two number-count models. The correlations
  between the number counts models are less significant.

\end{enumerate}

%% file: acks.tex
\section*{Acknowledgments}
\label{sec:acks}



SC  acknowledge support from the Claude Leon Foundation, 
SC and MGS acknowledge the South African Square Kilometre Array Project, 
the South African National Research Foundation.  
JZ gratefully acknowledges a South Africa National Research Foundation Square
Kilometre Array Research Fellowship. 
The computations described in this work were carried out on the African Research Cloud. 
We would especially like to thank Timothy Carr for his efficient technical support.
We also acknowledge Tessa Vernstrom for valuable comments and discussions.

%% file: pdstack_v6.bbl
\begin{thebibliography}{}
\makeatletter
\relax
\def\mn@urlcharsother{\let\do\@makeother \do\$\do\&\do\#\do\^\do\_\do\%\do\~}
\def\mn@doi{\begingroup\mn@urlcharsother \@ifnextchar [ {\mn@doi@}
  {\mn@doi@[]}}
\def\mn@doi@[#1]#2{\def\@tempa{#1}\ifx\@tempa\@empty \href
  {http://dx.doi.org/#2} {doi:#2}\else \href {http://dx.doi.org/#2} {#1}\fi
  \endgroup}
\def\mn@eprint#1#2{\mn@eprint@#1:#2::\@nil}
\def\mn@eprint@arXiv#1{\href {http://arxiv.org/abs/#1} {{\tt arXiv:#1}}}
\def\mn@eprint@dblp#1{\href {http://dblp.uni-trier.de/rec/bibtex/#1.xml}
  {dblp:#1}}
\def\mn@eprint@#1:#2:#3:#4\@nil{\def\@tempa {#1}\def\@tempb {#2}\def\@tempc
  {#3}\ifx \@tempc \@empty \let \@tempc \@tempb \let \@tempb \@tempa \fi \ifx
  \@tempb \@empty \def\@tempb {arXiv}\fi \@ifundefined
  {mn@eprint@\@tempb}{\@tempb:\@tempc}{\expandafter \expandafter \csname
  mn@eprint@\@tempb\endcsname \expandafter{\@tempc}}}

\bibitem[\protect\citeauthoryear{{Bondi} et~al.}{{Bondi}
  et~al.}{2003}]{bondi2003}
{Bondi} M.,  et~al., 2003, \mn@doi [\aap] {10.1051/0004-6361:20030382}, \href
  {http://adsabs.harvard.edu/abs/2003A%26A...403..857B} {403, 857}

\bibitem[\protect\citeauthoryear{{Bridges}, {Feroz}, {Hobson}  \&
  {Lasenby}}{{Bridges} et~al.}{2009}]{optimal-binning}
{Bridges} M.,  {Feroz} F.,  {Hobson} M.~P.,   {Lasenby} A.~N.,  2009, \mn@doi
  [\mnras] {10.1111/j.1365-2966.2009.15525.x}, \href
  {http://adsabs.harvard.edu/abs/2009MNRAS.400.1075B} {400, 1075}

\bibitem[\protect\citeauthoryear{{Buchner} et~al.,}{{Buchner}
  et~al.}{2014}]{pymultinest}
{Buchner} J.,  et~al., 2014, \mn@doi [\aap] {10.1051/0004-6361/201322971},
  \href {http://saaoads.chpc.ac.za/abs/2014A%26A...564A.125B} {564, A125}

\bibitem[\protect\citeauthoryear{{Condon}}{{Condon}}{1974}]{condon74}
{Condon} J.~J.,  1974, \apj, \href
  {http://ukads.nottingham.ac.uk/abs/1974ApJ...188..279C} {188, 279}

\bibitem[\protect\citeauthoryear{{Condon} et~al.,}{{Condon}
  et~al.}{2012a}]{condon2012}
{Condon} J.~J.,  et~al., 2012a, \mn@doi [\apj] {10.1088/0004-637X/758/1/23},
  \href {http://adsabs.harvard.edu/abs/2012ApJ...758...23C} {758, 23}

\bibitem[\protect\citeauthoryear{{Condon} et~al.,}{{Condon}
  et~al.}{2012b}]{2012ApJ...758...23C}
{Condon} J.~J.,  et~al., 2012b, \mn@doi [\apj] {10.1088/0004-637X/758/1/23},
  \href {http://adsabs.harvard.edu/abs/2012ApJ...758...23C} {758, 23}

\bibitem[\protect\citeauthoryear{{Elson}, {Blyth}  \& {Baker}}{{Elson}
  et~al.}{2016}]{ed2016}
{Elson} E.~C.,  {Blyth} S.~L.,   {Baker} A.~J.,  2016, \mn@doi [\mnras]
  {10.1093/mnras/stw1291}, \href
  {http://adsabs.harvard.edu/abs/2016MNRAS.460.4366E} {460, 4366}

\bibitem[\protect\citeauthoryear{{Feroz} \& {Hobson}}{{Feroz} \&
  {Hobson}}{2008}]{feroz08}
{Feroz} F.,  {Hobson} M.~P.,  2008, \mn@doi [\mnras]
  {10.1111/j.1365-2966.2007.12353.x}, \href
  {http://ukads.nottingham.ac.uk/abs/2008MNRAS.384..449F} {384, 449}

\bibitem[\protect\citeauthoryear{{Feroz}, {Hobson}  \& {Bridges}}{{Feroz}
  et~al.}{2009}]{feroz09}
{Feroz} F.,  {Hobson} M.~P.,   {Bridges} M.,  2009, \mn@doi [\mnras]
  {10.1111/j.1365-2966.2009.14548.x}, \href
  {http://adsabs.harvard.edu/abs/2009MNRAS.398.1601F} {398, 1601}

\bibitem[\protect\citeauthoryear{{Franzen} et~al.,}{{Franzen}
  et~al.}{2016}]{franzen2016}
{Franzen} T.~M.~O.,  et~al., 2016, \mn@doi [\mnras] {10.1093/mnras/stw823},
  \href {http://adsabs.harvard.edu/abs/2016MNRAS.459.3314F} {459, 3314}

\bibitem[\protect\citeauthoryear{{Heald} et~al.,}{{Heald}
  et~al.}{2015}]{2015A&A...582A.123H}
{Heald} G.~H.,  et~al., 2015, \mn@doi [\aap] {10.1051/0004-6361/201425210},
  \href {http://adsabs.harvard.edu/abs/2015A%26A...582A.123H} {582, A123}

\bibitem[\protect\citeauthoryear{{Hewish}}{{Hewish}}{1961}]{hewish61}
{Hewish} A.,  1961, \mnras, \href
  {http://adsabs.harvard.edu/abs/1961MNRAS.123..167H} {123, 167}

\bibitem[\protect\citeauthoryear{{Jarvis} et~al.}{{Jarvis}
  et~al.}{2013}]{jarvis2013}
{Jarvis} M.~J.,  et~al., 2013, \mn@doi [\mnras] {10.1093/mnras/sts118}, \href
  {http://adsabs.harvard.edu/abs/2013MNRAS.428.1281J} {428, 1281}

\bibitem[\protect\citeauthoryear{{Jarvis} et~al.}{{Jarvis}
  et~al.}{2017}]{mightee2017}
{Jarvis} M.~J.,  et~al., 2017, preprint, \href
  {http://adsabs.harvard.edu/abs/2017arXiv170901901J} {} (\mn@eprint {arXiv}
  {1709.01901})

\bibitem[\protect\citeauthoryear{{Longair}}{{Longair}}{1966}]{Longair1966}
{Longair} M.~S.,  1966, Mon. Not. R. Astron. Soc., \href
  {http://adsabs.harvard.edu/abs/1966MNRAS.133..421L} {133, 421}

\bibitem[\protect\citeauthoryear{{Mitchell-Wynne}, {Santos}, {Afonso}  \&
  {Jarvis}}{{Mitchell-Wynne} et~al.}{2014}]{ketron2014}
{Mitchell-Wynne} K.,  {Santos} M.~G.,  {Afonso} J.,   {Jarvis} M.~J.,  2014,
  \mn@doi [\mnras] {10.1093/mnras/stt2035}, \href
  {http://adsabs.harvard.edu/abs/2014MNRAS.437.2270M} {437, 2270}

\bibitem[\protect\citeauthoryear{{Ryle}}{{Ryle}}{1961}]{Ryle1961}
{Ryle} M.,  1961, \mn@doi [Nature] {10.1038/190852a0}, \href
  {http://adsabs.harvard.edu/abs/1961Natur.190..852R} {190, 852}

\bibitem[\protect\citeauthoryear{{Ryle} \& {Clarke}}{{Ryle} \&
  {Clarke}}{1961}]{RyleClarke}
{Ryle} M.,  {Clarke} R.~W.,  1961, Mon. Not. R. Astron. Soc., 122, 349

\bibitem[\protect\citeauthoryear{{Scheuer}}{{Scheuer}}{1957}]{scheuer57}
{Scheuer} P.~A.~G.,  1957, in Proceedings of the Cambridge Philisophical
  Society. pp 764--773

\bibitem[\protect\citeauthoryear{{Skilling}}{{Skilling}}{2004}]{skilling04}
{Skilling} J.,  2004, in {Fischer} R.,  {Preuss} R.,   {Toussaint} U.~V.,  eds,
  American Institute of Physics Conference Series. pp 395--405,
  \mn@doi{10.1063/1.1835238}

\bibitem[\protect\citeauthoryear{{Vernstrom} et~al.,}{{Vernstrom}
  et~al.}{2014}]{vernstrom2014}
{Vernstrom} T.,  et~al., 2014, \mn@doi [\mnras] {10.1093/mnras/stu470}, \href
  {http://adsabs.harvard.edu/abs/2014MNRAS.440.2791V} {440, 2791}

\bibitem[\protect\citeauthoryear{{Wilman} et~al.}{{Wilman}
  et~al.}{2008}]{wilman2008}
{Wilman} R.~J.,  et~al., 2008, \mn@doi [\mnras]
  {10.1111/j.1365-2966.2008.13486.x}, \href
  {http://adsabs.harvard.edu/abs/2008MNRAS.388.1335W} {388, 1335}

\bibitem[\protect\citeauthoryear{{Wilman}, {Jarvis}, {Mauch}, {Rawlings}  \&
  {Hickey}}{{Wilman} et~al.}{2010a}]{skads2}
{Wilman} R.~J.,  {Jarvis} M.~J.,  {Mauch} T.,  {Rawlings} S.,   {Hickey} S.,
  2010a, \mn@doi [\mnras] {10.1111/j.1365-2966.2010.16453.x}, \href
  {http://adsabs.harvard.edu/abs/2010MNRAS.405..447W} {405, 447}

\bibitem[\protect\citeauthoryear{{Wilman}, {Jarvis}, {Mauch}, {Rawlings}  \&
  {Hickey}}{{Wilman} et~al.}{2010b}]{wilman2010}
{Wilman} R.~J.,  {Jarvis} M.~J.,  {Mauch} T.,  {Rawlings} S.,   {Hickey} S.,
  2010b, \mn@doi [\mnras] {10.1111/j.1365-2966.2010.16453.x}, \href
  {http://adsabs.harvard.edu/abs/2010MNRAS.405..447W} {405, 447}

\bibitem[\protect\citeauthoryear{{Zwart} et~al.,}{{Zwart}
  et~al.}{2015a}]{jz-skasci}
{Zwart} J.,  et~al., 2015a, Advancing Astrophysics with the Square Kilometre
  Array (AASKA14), \href {http://adsabs.harvard.edu/abs/2015aska.confE.172Z}
  {p.~172}

\bibitem[\protect\citeauthoryear{{Zwart}, {Santos}  \& {Jarvis}}{{Zwart}
  et~al.}{2015b}]{farbeyond}
{Zwart} J.~T.~L.,  {Santos} M.,   {Jarvis} M.~J.,  2015b, \mn@doi [\mnras]
  {10.1093/mnras/stv1716}, \href
  {http://adsabs.harvard.edu/abs/2015MNRAS.453.1740Z} {453, 1740}

\makeatother
\end{thebibliography}
